\makeatletter
\def\input@path{{tex/latex/revtex/}}
\makeatother
\documentclass[%
 reprint,
superscriptaddress,
 amsmath,amssymb,
 aps,
 pra,
]{revtex4-2}

  \usepackage[pdftex]{graphicx}
  \DeclareGraphicsExtensions{.pdf,.jpeg,.png}

\usepackage{amsmath}

\usepackage{algorithmic}

\usepackage{array}

\usepackage[font=normalsize,labelfont=sf,textfont=sf]{subcaption}

\usepackage[version=4]{mhchem}
\usepackage{eurosym}
\usepackage[activate={true,nocompatibility},final,tracking=true,kerning=true,spacing=true,factor=1100,stretch=10,shrink=10]{microtype}
\usepackage[free-standing-units]{siunitx}
\usepackage[normalem]{ulem}
\usepackage{soul}
\usepackage[colorlinks]{hyperref}
\usepackage{url}
\usepackage{cleveref}

\newif\ifuselaserschematic
\uselaserschematictrue 

\newcommand\pythonimagescaling{0.45}

\begin{document}
\title{Sub-surface modifications in silicon with ultra-short pulsed lasers above 2 microns}

\author{Roland~A.~Richter}
\affiliation{Department of Physics, NTNU, Norwegian University of Science and Technology, Trondheim, Norway}
\author{Nikolai~Tolstik}
\affiliation{Department of Physics, NTNU, Norwegian University of Science and Technology, Trondheim, Norway}
\affiliation{ATLA Lasers AS}
\author{Sebastien~Rigaud}
\affiliation{Department of Physics, NTNU, Norwegian University of Science and Technology, Trondheim, Norway}
\author{Paul~Dalla~Valle}
\affiliation{Department of Physics, NTNU, Norwegian University of Science and Technology, Trondheim, Norway}
\author{Andreas~Erbe}
\affiliation{Department of Materials Science and Engineering, NTNU, Norwegian University of Science and Technology, Trondheim, Norway}
\author{Petra~Ebbinghaus}
\affiliation{Max-Planck-Institut f\"ur Eisenforschung GmbH, D\"usseldorf, Germany}
\author{Ignas~Astrauskas}
\affiliation{Photonics Institute, Vienna University of Technology, Vienna, Austria}
\author{Vladimir~Kalashnikov}
\affiliation{Photonics Institute, Vienna University of Technology, Vienna, Austria}
\affiliation{Dipartimento di Ingegneria dell'Informazione, Sapienza University of Rome, Italy}
\author{Evgeni~Sorokin}
\affiliation{Photonics Institute, Vienna University of Technology, Vienna, Austria}
\author{Irina~T.~Sorokina}
\email{sorokina@ntnu.no}
\affiliation{Department of Physics, NTNU, Norwegian University of Science and Technology, Trondheim, Norway}
\affiliation{ATLA Lasers AS}
%

\begin{abstract}
Nonlinear optical phenomena in silicon such as self-focusing and multi-photon absorption are strongly dependent on the wavelength, energy and duration of the exciting pulse. Thus, a pronounced wavelength dependence of the sub-surface modifications with ultra-short pulsed lasers exists, especially for wavelengths $>$2~{\micro\meter}. This wavelength dependence is investigated for wavelengths in the range of \SIrange{1950}{2400}{\nano\meter}, at a pulse duration between \SIrange{0.5}{10}{\pico\second} and the pulse energy varying from \SI{1}{\micro\joule} to \SI{1}{\milli\joule}. 
Numerical and experimental analyses have been performed on both the surface and sub-surface of Si wafers processed with fibre-based lasers built in-house that operate in this wavelength range. The results have been compared to the literature data at $ \SI{1550}{\nano\meter} $. The analysis carried out has shown that due to a dip in the nonlinear absorption spectrum and a peak in the spectrum of the third-order non-linearity, the wavelengths between \SIrange{2000}{2200}{\nano\meter} are more favourable for creating sub-surface modifications in silicon. This is the case even though those wavelengths do not allow as tight a focusing as those at \SI{1550}{\nano\meter} in the linear regime. This problem is compensated by an increased self-focusing due to the nonlinear Kerr-effect around \SI{2100}{\nano\meter} at high light intensities, characteristic for ultra-short pulses.
\end{abstract}
\maketitle



\section{INTRODUCTION AND STATE-OF-THE-ART}
\label{sec:IntroSotA}
Silicon wafers are currently largely separated from the bulk mono-crystalline Si-block using thin diamond saws, which introduces a loss of material of up to $ \SI{50}{\percent} $ \cite{Dross2012}. For extremely thin wafers with a thickness $ \leq\SI{100}{\micro\meter} $ the loss increases to \SI{70}{\percent}\cite{Dross2012}. Thus, alternative methods for wafer separation are in development, such as epitaxial Si lift-off, stress-induced spalling, and smart-cut \cite{Bruel1995,Lee2018}. This latter technique employs the fact that by the introduction of defects in a target layer below the Si wafer surface this layer will be weakened, allowing the wafer itself to be removed. There are multiple techniques for introducing such defects into bulk Si. The expansion coefficient difference for two materials can be used \cite{Lee2018_KF}, without adding modifications inside the crystal. However, the crystal itself can be modified internally by the use of a layer of porous Si\cite{Kajari-Schroder2013}, the use of high-energy protons \cite{Lee2018} or the use of two-photon absorption of laser pulses at \SI{1550}{\nano\meter}\cite{Verburg}.

Even though much work has been conducted at common wavelengths such as \SI{1064}{\nano\meter} and \SI{1550}{\nano\meter}, both experimentally and numerically \cite{Ramer2014,doi:10.1063/1.331020,Verburg2014,Richter:18,Zavedeev2016,Kumagai2007,Korfiatis2007,Gan2011} only a small amount of work has been done at longer wavelengths, including \SI{1965}{\nano\meter}\cite{Richter:18}, \SI{1970}{\nano\meter}\cite{Chambonneau2019} and \SI{2300}{\nano\meter}\cite{Nejadmalayeri2005, Richter:18}. Chambonneau et al. \cite{Chambonneau2019} used a \ce{Tm}-fiber laser for generating pulses with a duration of \SI{400}{\pico\second}, which then were compressed outside of the fiber laser. Furthermore, they investigated the number of pulses necessary to modify silicon, depending on the pulse energy. In the latter work a femtosecond OPO (optical parametric oscillator) was used to produce modifications below a \ce{SiO2} layer, but not within bulk Si. 

In the first section of the present paper we numerically investigate the influence of the wavelength, pulse energy and pulse duration on the defect formation of silicon using both numerical simulations and experiments for wavelengths between \SIrange{1950}{2400}{\nano\meter}. The results obtained are compared  with the existing data for 
\SI{1550}{\nano\meter}, \SI{1970}{\nano\meter} and \SI{2300}{\nano\meter}. Thereby we can confirm the preliminary results we showed in \cite{Richter:18}, that an optimal wavelength range for silicon processing exists. 

On the experimental side, this paper is separated into two separate parts. In those parts several different lasers have been used to modify a Si-crystal both on the surface and within the bulk material by using the wavelengths longer than \SI{1950}{\nano\meter}. In order to optimize the process both in quality and speed, several parameters such as pulse energy, pulse duration and wavelength had to be optimized. It has been already demonstrated that shorter pulses provide better cut quality and precision compared to longer pulses \cite{Jeschke2002,Chichkov1996}, while lowering the ablation threshold. We have previously shown that longer wavelengths in the mid-infrared range between \SI{2000}{\nano\meter} and \SI{2200}{\nano\meter} are beneficial for processing of bulk Si \cite{Richter:18}. This is due to the combination of lower non-linear absorption \cite{Lin2007} and higher self-focusing \cite{Wang2013}. 
Therefore, we present results showing the creation of sub-surface modifications within silicon by using pulses at nano-joule levels at two different wavelengths (\SI{1965}{\nano\meter} and \SI{2350}{\nano\meter}) in the first sub-part of the experimental section of this paper. Afterwards, we present the generation of sub-surface modifications in silicon using micro-joule level pulses at a wavelength of \SI{2090}{\nano\meter} in the second sub-part.

In summary we present here for to our knowledge the first time the results of both numerical and experimental studies above \SI{1550}{\nano\meter} using fiber based lasers with a wavelength of \SI{1965}{\nano\meter}, \SI{2090}{\nano\meter} and \SI{2350}{\nano\meter} at different pulse energies. The results show not only that silicon processing is favourable at longer wavelengths (as it was shown before in previous studies \cite{Richter:18,Chambonneau2019}), but also that an optimum wavelength range for modifying silicon exists. This is based on the wavelength dependencies of the non-linear refractive index $n_2$ and the multi-photon absorption coefficients $\beta$ and $\gamma$ for two- and three-photon absorption, respectively.
\section{METHODOLOGY}
\label{sec:Methodology}
The spectrum of the non-linear multi-photon absorption calculated (in our case) as a sum of the two-photon- and three-photon-absorption has a ``dip'' within the wavelength range we target (between \SIrange{2000}{2200}{\nano\meter}), as shown in \cite{Richter:18}. The values for two-photon-absorption decrease above $ \approx\SI{1800}{\nano\meter} $, while the three-photon-absorption is noticeably rising above \SI{2300}{\nano\meter}, thus leaving a gap in between\cite{Richter:18,Zavedeev2016}.

The Nonlinear Figure of Merit (NFoM) characterizing the relative merit of the Kerr nonlinearity coefficient versus the two-photon nonlinearity is usually defined as the ratio between both values, divided by the optical wavelength in vacuum \cite{Mizrahi1989}. Extending this expression to multi-photon absorption results in the NFoM that is written in the following form \cite{Wang2013}:
\begin{equation}\label{equ:FOM-Equation}
	NFOM=\sum_K\frac{n_2(\lambda)}{\lambda\beta^{(K)}I^{(K-1)}}
\end{equation}
with $ \lambda $ the wavelength in vacuum, $ n_2(\lambda) $ the wavelength-dependent refractive index, $ I $ the field intensity and $ \beta^{(K)} $ the coefficient for $ K $-photon absorption.

Based on \Cref{equ:FOM-Equation}, an optimal wavelength range for generating sub-surface modifications could be estimated to be in the range of \SIrange{2000}{2300}{\nano\meter} \cite{Richter:18}. The exact wavelength depends on the exact value of the material parameters involved, such as the value for the wavelength-dependent Kerr-nonlinearity $n_2(\lambda)$, which again depends on the material properties (e.g. single-crystal versus polycrystalline material, n-doped versus p-doped). This explains a large range of the reported values of silicon nonlinearities, including measurement errors up to \SI{30}{\percent} \cite{Lin2007,Wang2013}. Thus, one can only predict a certain wavelength range, not an exact value of the optimal wavelength for generating sub-surface modifications.

To model the propagation of the pulse through the material, we use the non-linear Schr\"{o}dinger equation \cite{Zavedeev2016}
\begin{equation}\label{equ:NLS-equation}
\begin{split}
\partial_zE&=\frac{i}{2k_0}\Delta E+\frac{ik_0n_2}{n_0}\vert E\vert^2E\\&\quad-\frac{1}{2}\sum_K\beta^{(K)}\vert E\vert^{2K-2}E\\&\quad-\frac{\sigma}{2}\left(1+i\omega_0\tau_c\right)NE
\end{split}
\end{equation}

\Cref{equ:NLS-equation} describes the propagation of a time ($ t $)--dependent and transversely inhomogeneous laser field $ E $ through bulk Si along the $ z $-axis. The diffraction is taken into account by the transverse Laplacian, where $ k_0 $ is the wavenumber. The nonlinear effects of self-phase modulation and self-focusing are described by the term proportional to $ n_2 $, and $ \tau_c=\SI{3.5}{\femto\second} $ is the free-carriers momentum scattering time \cite{Zavedeev2016}. The non-linear multi-photon absorption for $ K $-photon-absorption $ \left(\beta^{(K)}\right) $, the absorption caused by free carriers with the concentration $ N $, and the inverse bremsstrahlung absorption $ \sigma $ are taken into account, as well. Since the investigated wavelengths are in the range where the single-photon-absorption and higher-photon-absorption could be neglected \cite{Gholami2011}, we have considered the action of only two- and three-photon absorption. The numerical simulation was done while using support functions from the library deal.II \cite{dealII91}.

The paraxial approximation introduces several simplifications into the beam propagation equation, but is only valid until a certain beam diameter \cite{Nemoto1990}. Thus, if the beam is focused too strongly because of the optical Kerr-effect, \Cref{equ:NLS-equation} which uses that approximation loses its' validity. Deviations from paraxiality also have to be taken into account when simulating lenses with extremely high NA. 

There are different approaches for calculating the response of the material if irradiated with ultra-short pulses, which are extensively discussed in \cite{Rethfeld2017}. Examples include the two-temperature-model (TTM) and its extension, the three-temperature model \cite{Lee2006}, a hydrodynamic approach and molecular dynamics (MD) simulations. Furthermore, hybrid models such as a combination of a TTM and a MD are suggested \cite{Rethfeld2017}. In order to simplify calculations, in this study we take into account only the generated carrier density. This value can be calculated by using \cite{Zavedeev2016}: 
\begin{equation}\label{equ:CarrierDensityGeneration}
\partial_tN=\sum_K\frac{\beta^{(K)}\vert E\vert^{2K}}{K\hbar\omega_0}
\end{equation}
where $ \vert E\vert^2\equiv I $.

\section{NUMERICAL SIMULATIONS}
\label{sec:NumSimulations}
An initial rough comparison of the action produced on silicon by ultra-short pulses at different wavelengths can be done using \Cref{equ:NLS-equation}, by comparing the transported energy towards the focal spot. The results of such simulations for the wavelengths \SI{1550}{\nano\meter}, \SI{1950}{\nano\meter}, \SI{2150}{\nano\meter} and \SI{2350}{\nano\meter} with a pulse energy of \SI{1}{\micro\joule}, \SI{10}{\micro\joule} and \SI{100}{\micro\joule} for two different values of $ n_2 $ \cite{Lin2007,Wang2013} are shown in \Cref{fig:energy1u10u100u}. The results denoted with ``high $ n_2 $-values'' and ``low $ n_2 $-values'' use $n_2$-values taken from \cite{Wang2013} and \cite{Lin2007}, respectively. The pulse duration is set to \SI{5}{\pico\second}, to correlate as close as possible with the experiments presented in \Cref{sec:ExpDiscussion}. In these experiments, a large range of different lenses and focal depths was used. To limit the presented data to a reasonable amount, while still being able to demonstrate representative results, in the simulations we selected a lens with a focal length of \SI{8}{\milli\meter}.The beam radius was adjusted to have a comparable focal spot area for linear focusing, with a beam radius of \SI{1}{\milli\meter} for the wavelength of \SI{1550}{\nano\meter}.

\begin{figure}[htpb]
    \centering
    \begin{subfigure}[t]{\pythonimagescaling\textwidth}
    \includegraphics[width=\textwidth]{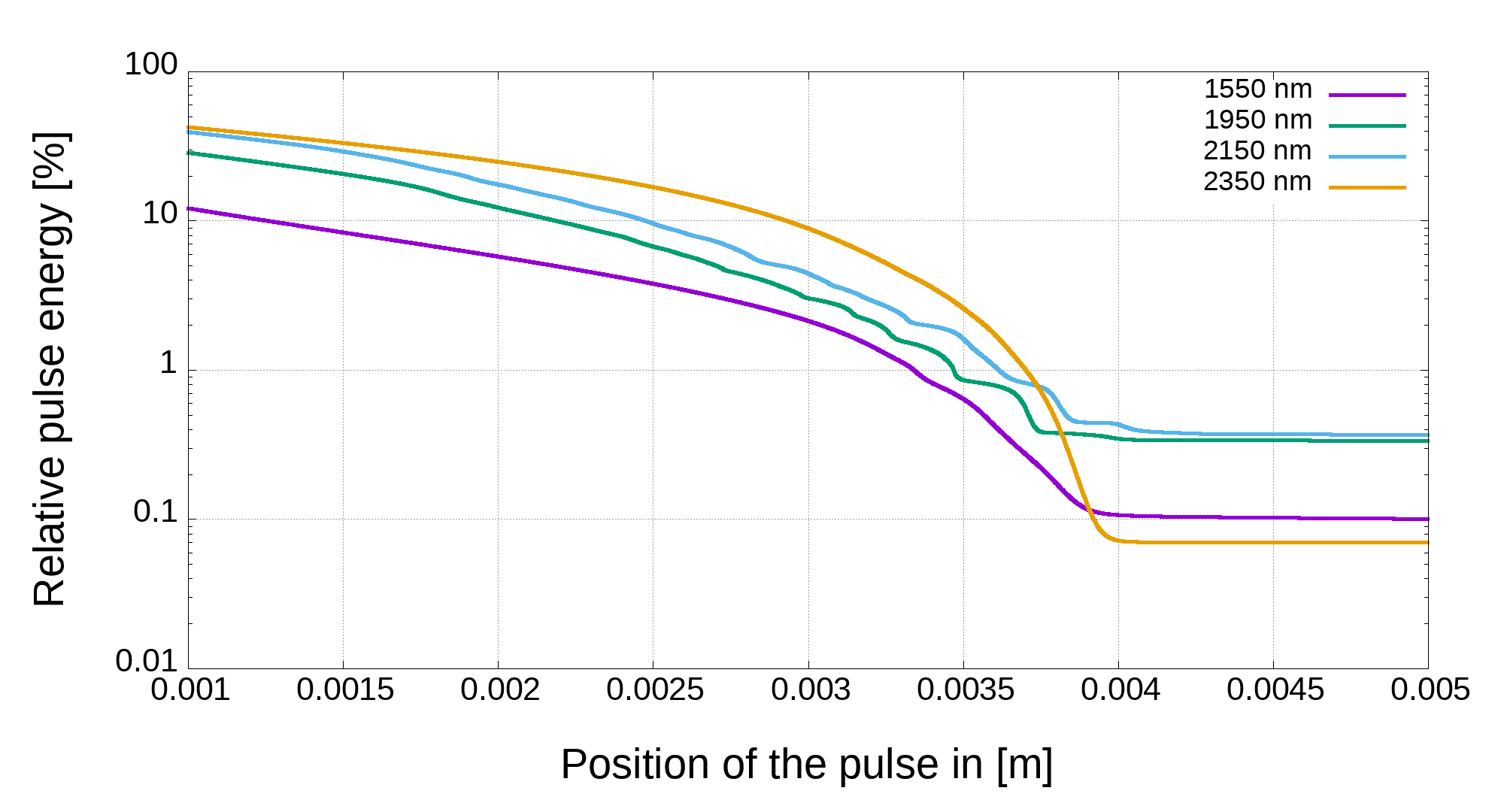}
    \caption{High $n_2$-values, \SI{1}{\micro\joule} pulse energy}
    \label{fig:highn21u}
    \end{subfigure}
    \hfill
    \begin{subfigure}[t]{\pythonimagescaling\textwidth}
    \includegraphics[width=\textwidth]{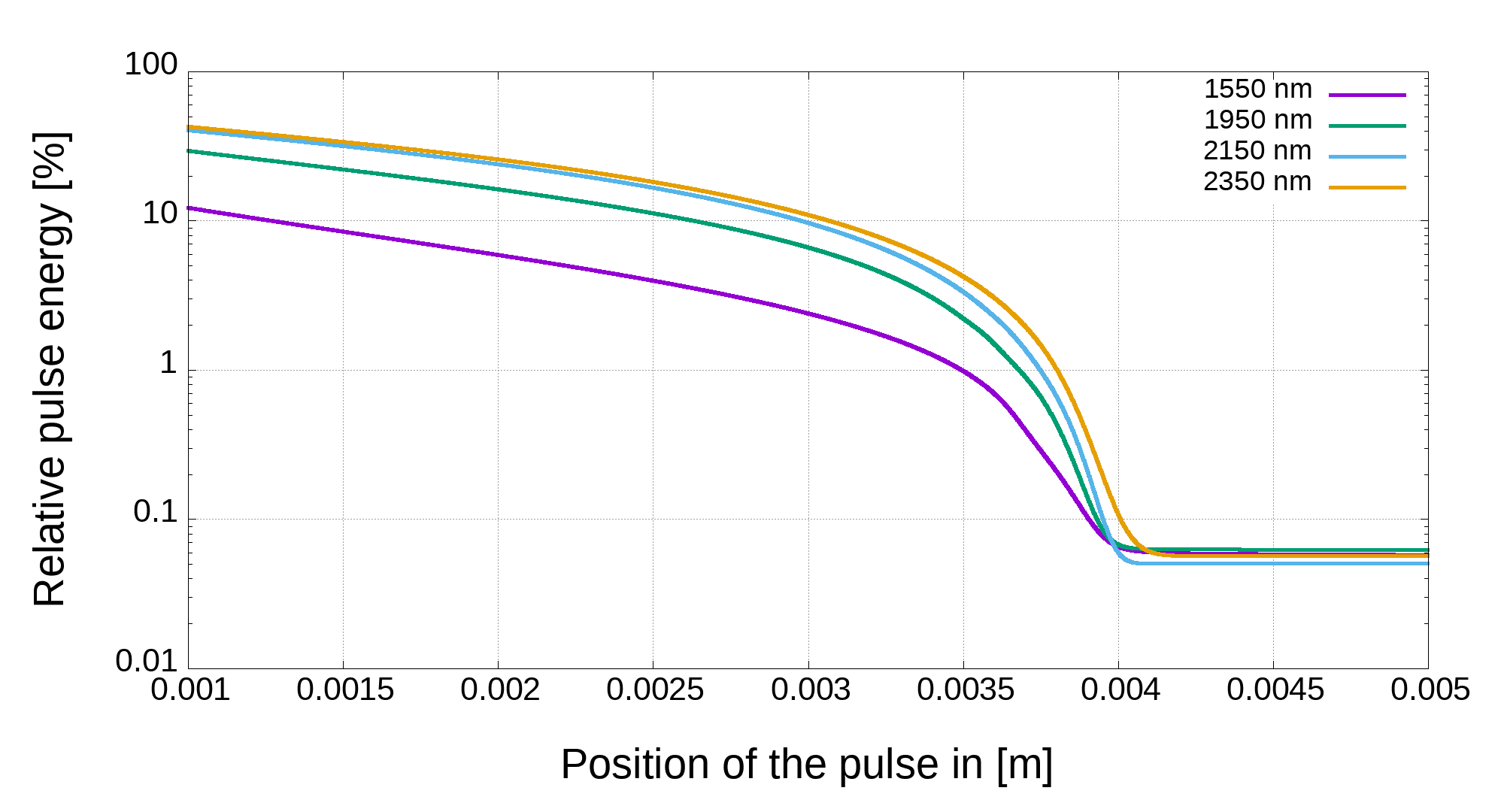}
    \caption{Low $n_2$-values, \SI{1}{\micro\joule} pulse energy}
    \label{fig:lown21u}
    \end{subfigure}
    \hfill
    \begin{subfigure}[t]{\pythonimagescaling\textwidth}
    \includegraphics[width=\textwidth]{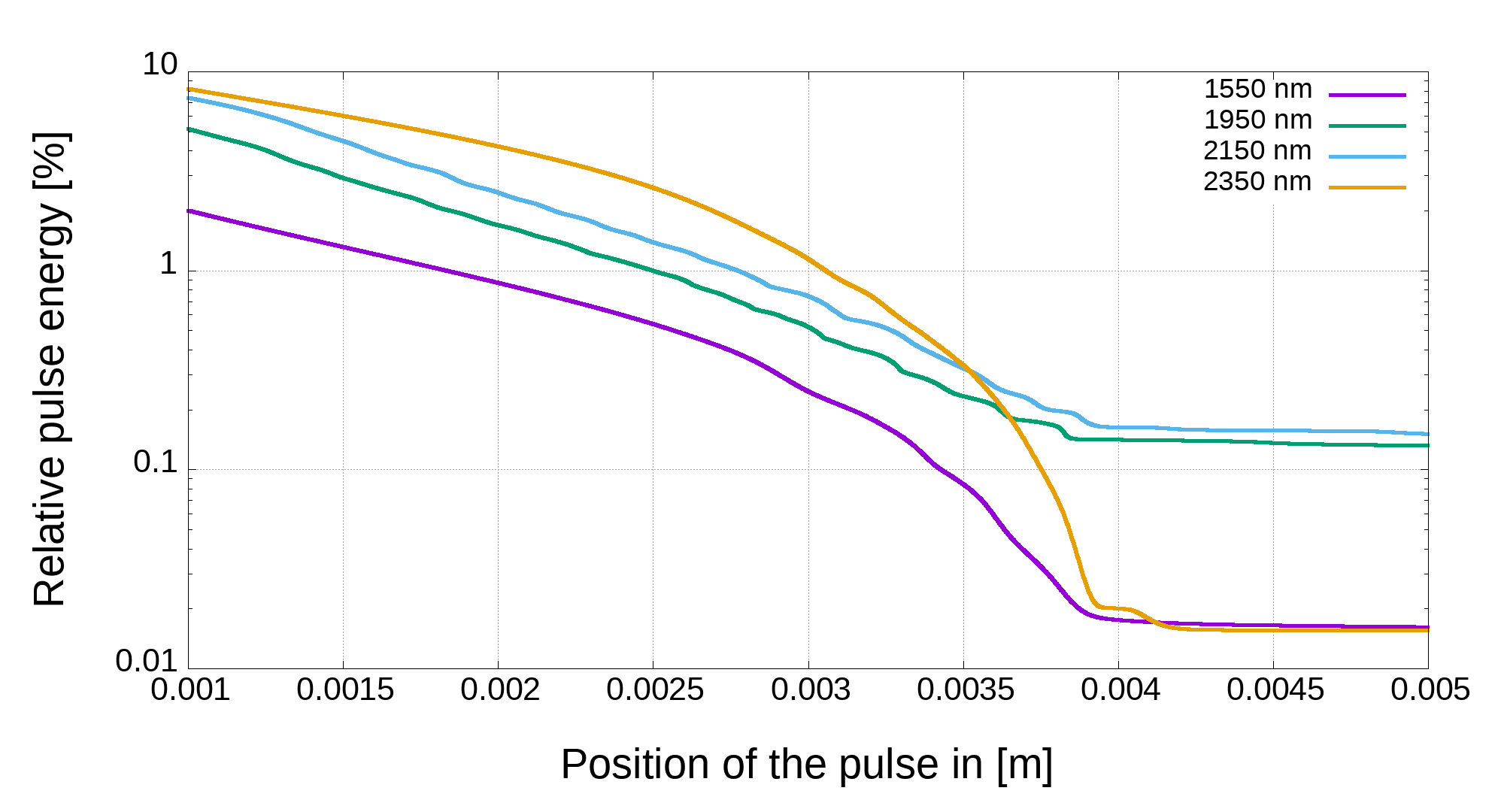}
    \caption{High $n_2$-values, \SI{10}{\micro\joule} pulse energy}
    \label{fig:highn210u}
    \end{subfigure}
    \hfill
    \begin{subfigure}[t]{\pythonimagescaling\textwidth}
    \includegraphics[width=\textwidth]{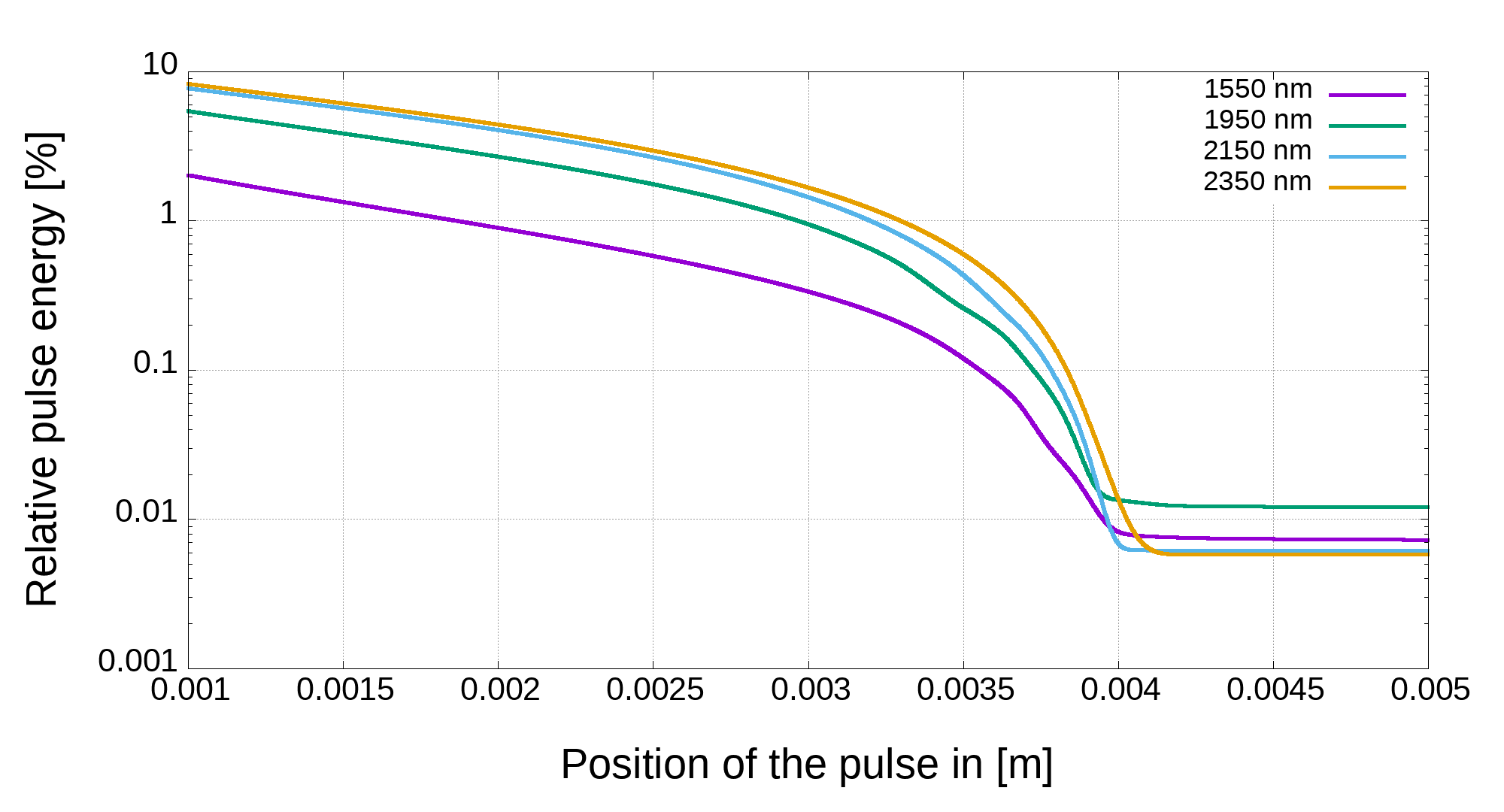}
    \caption{Low $n_2$-values, \SI{10}{\micro\joule} pulse energy}
    \label{fig:lown210u}
    \end{subfigure}
\end{figure}
\begin{figure}[htpb]
\ContinuedFloat
    \begin{subfigure}[t]{\pythonimagescaling\textwidth}
    \includegraphics[width=\textwidth]{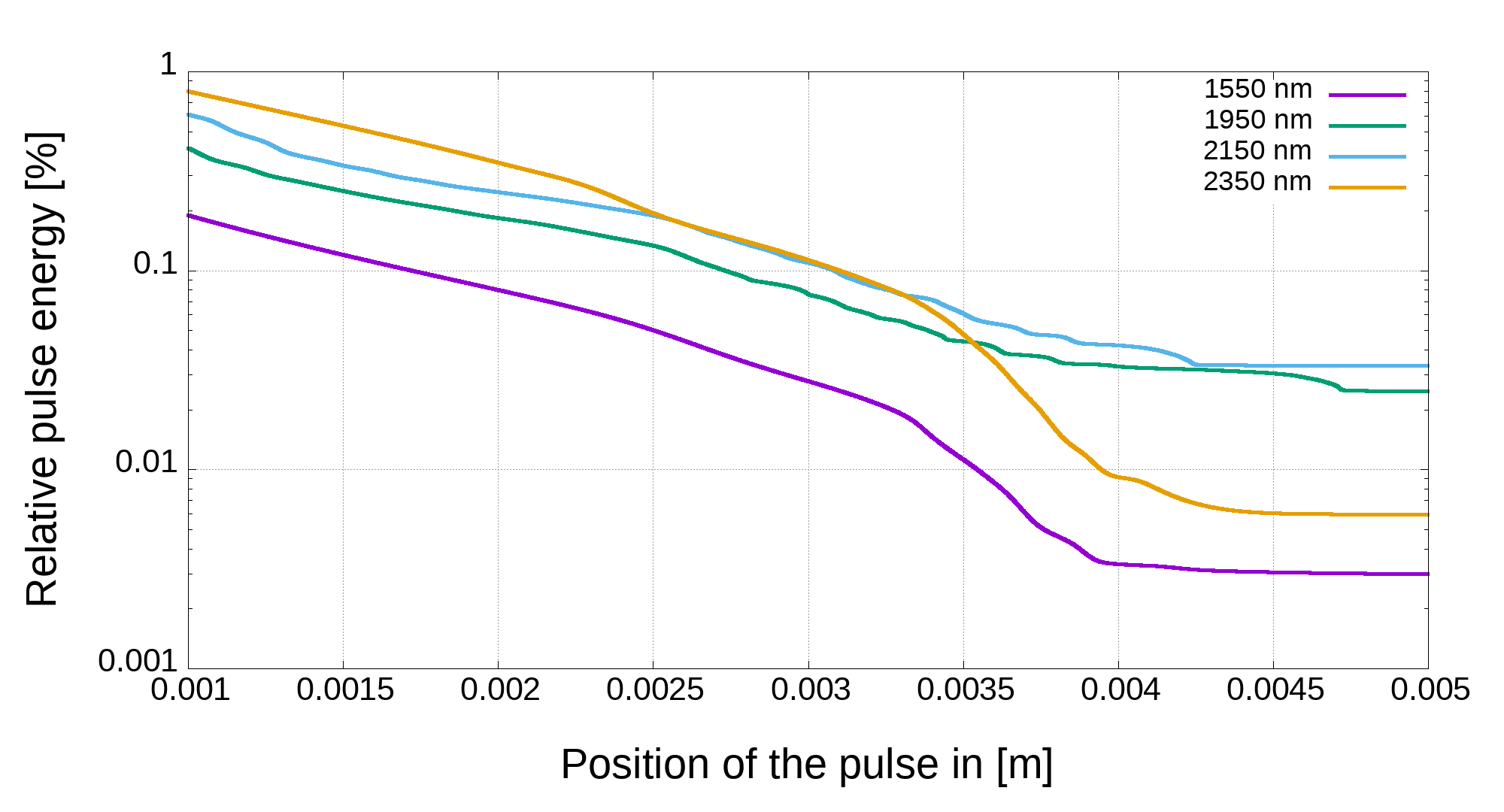}
    \caption{High $n_2$-values, \SI{100}{\micro\joule} pulse energy}
    \label{fig:highn2100u}
    \end{subfigure}
    \hfill
    \begin{subfigure}[t]{\pythonimagescaling\textwidth}
    \includegraphics[width=\textwidth]{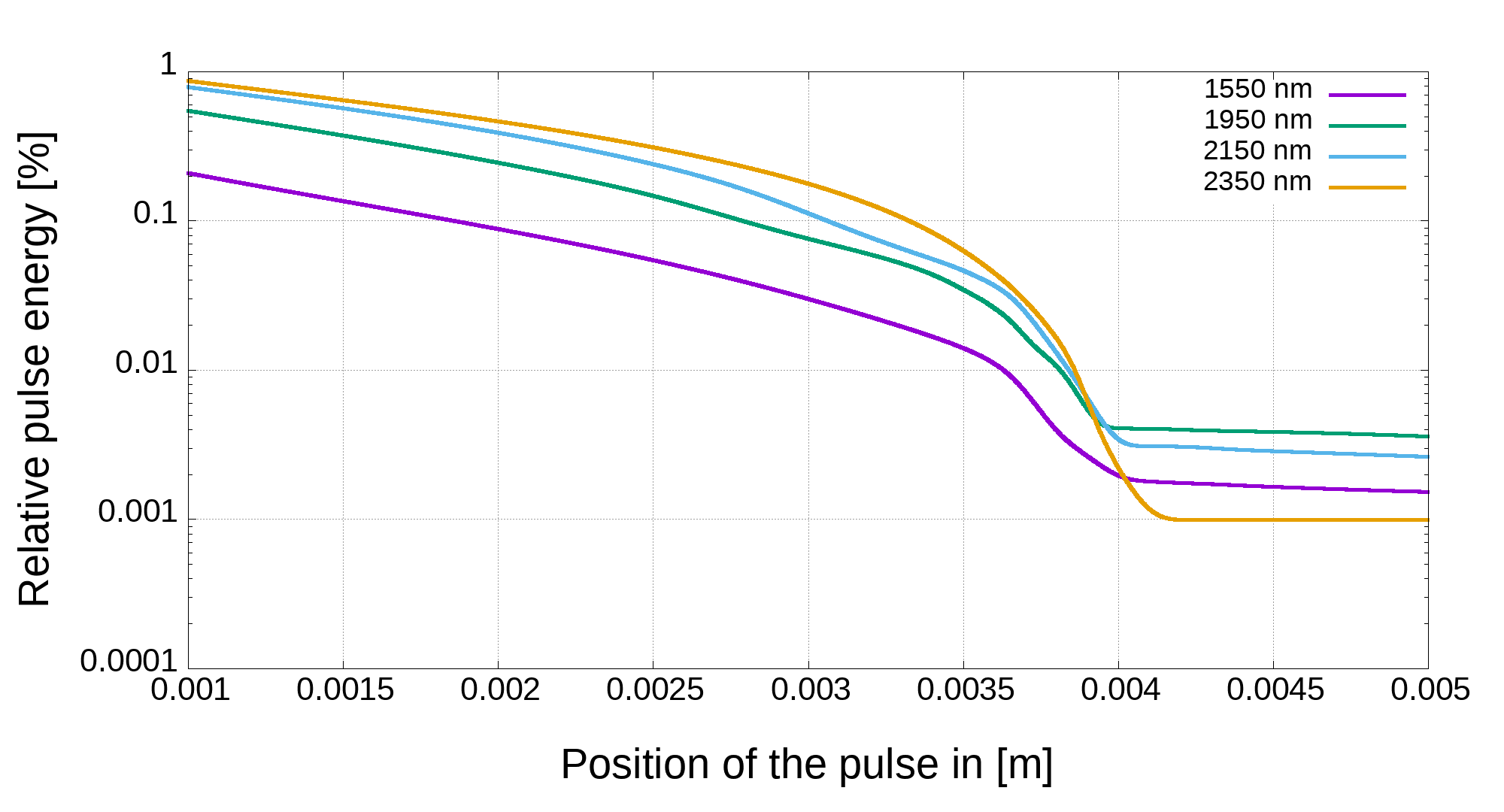}
    \caption{Low $n_2$-values, \SI{100}{\micro\joule} pulse energy}
    \label{fig:lown2100u}
    \end{subfigure}
    \caption{Energy of a pulse with different wavelengths and energies, depending on the position within the material. The focal length of the used lens is $f = \SI{8}{\milli\meter}$, with the focal point at $f_i = \SI{4}{\milli\meter}$ inside the material. The pulse duration is $\SI{5}{\pico\second}$, and the pulse energy is varied between \SI{1}{\micro\joule} and \SI{100}{\micro\joule}.}
    \label{fig:energy1u10u100u}
\end{figure}

\begin{figure}[htpb]
    \centering
    \begin{subfigure}[t]{\pythonimagescaling\textwidth}
    \includegraphics[width=\textwidth]{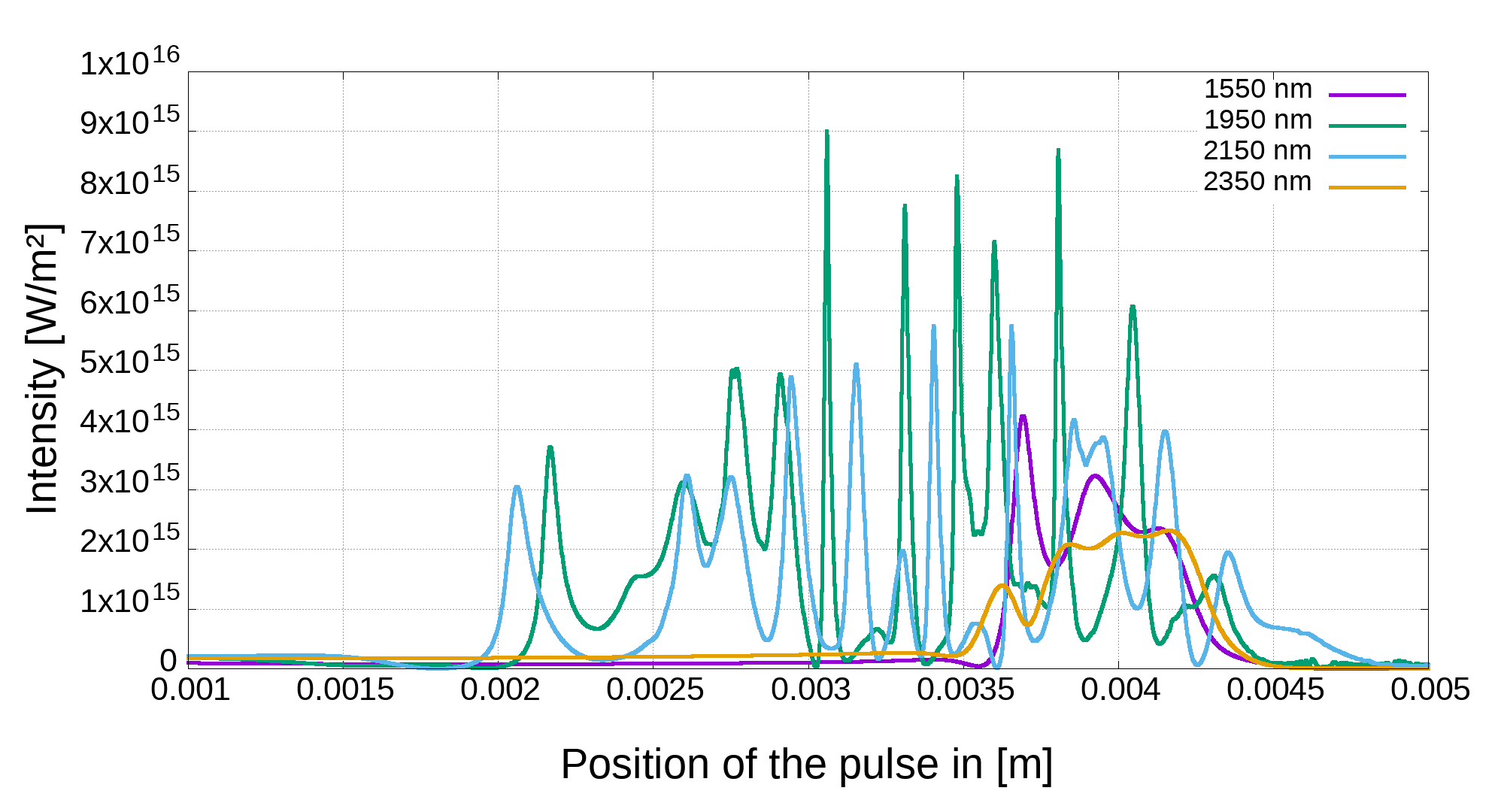}
    \caption{High $n_2$-values, \SI{1}{\micro\joule} pulse energy}
    \label{fig:highn21u_int}
    \end{subfigure}
    \hfill
    \begin{subfigure}[t]{\pythonimagescaling\textwidth}
    \includegraphics[width=\textwidth]{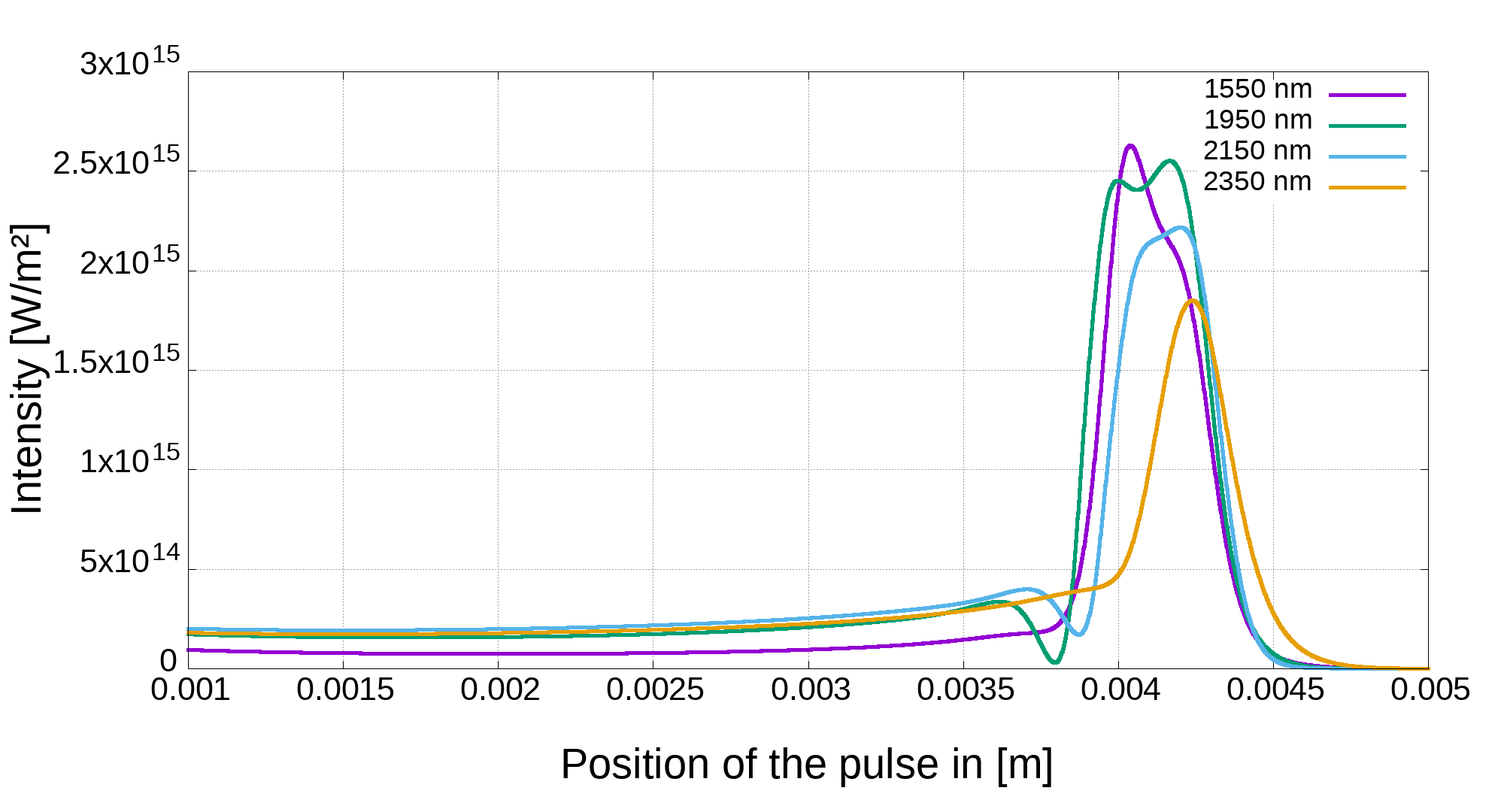}
    \caption{Low $n_2$-values, \SI{1}{\micro\joule} pulse energy}
    \label{fig:lown21u_int}
    \end{subfigure}
    \hfill
    \begin{subfigure}[t]{\pythonimagescaling\textwidth}
    \includegraphics[width=\textwidth]{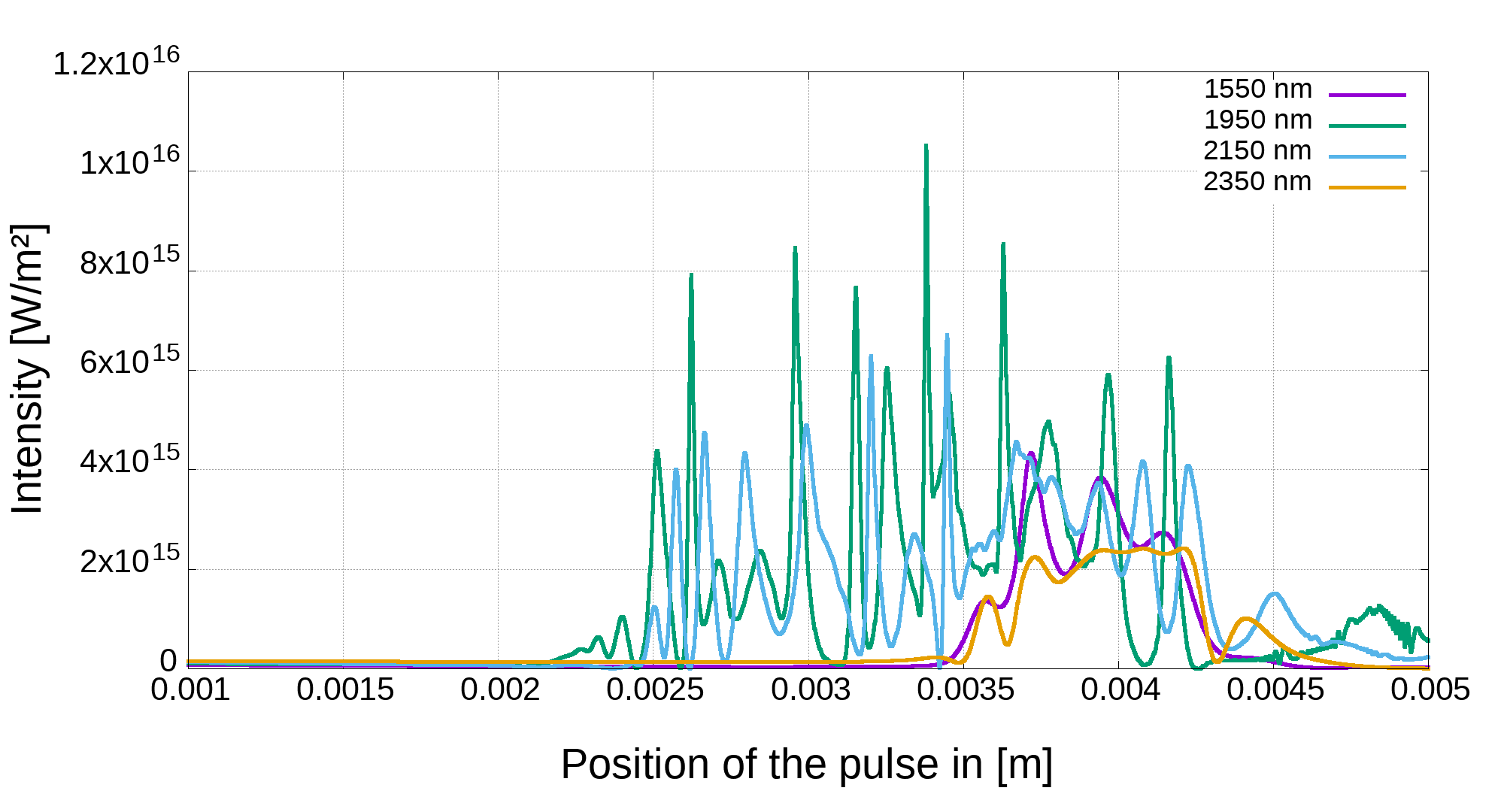}
    \caption{High $n_2$-values, \SI{10}{\micro\joule} pulse energy}
    \label{fig:highn210u_int}
    \end{subfigure}
    \hfill
    \begin{subfigure}[t]{\pythonimagescaling\textwidth}
    \includegraphics[width=\textwidth]{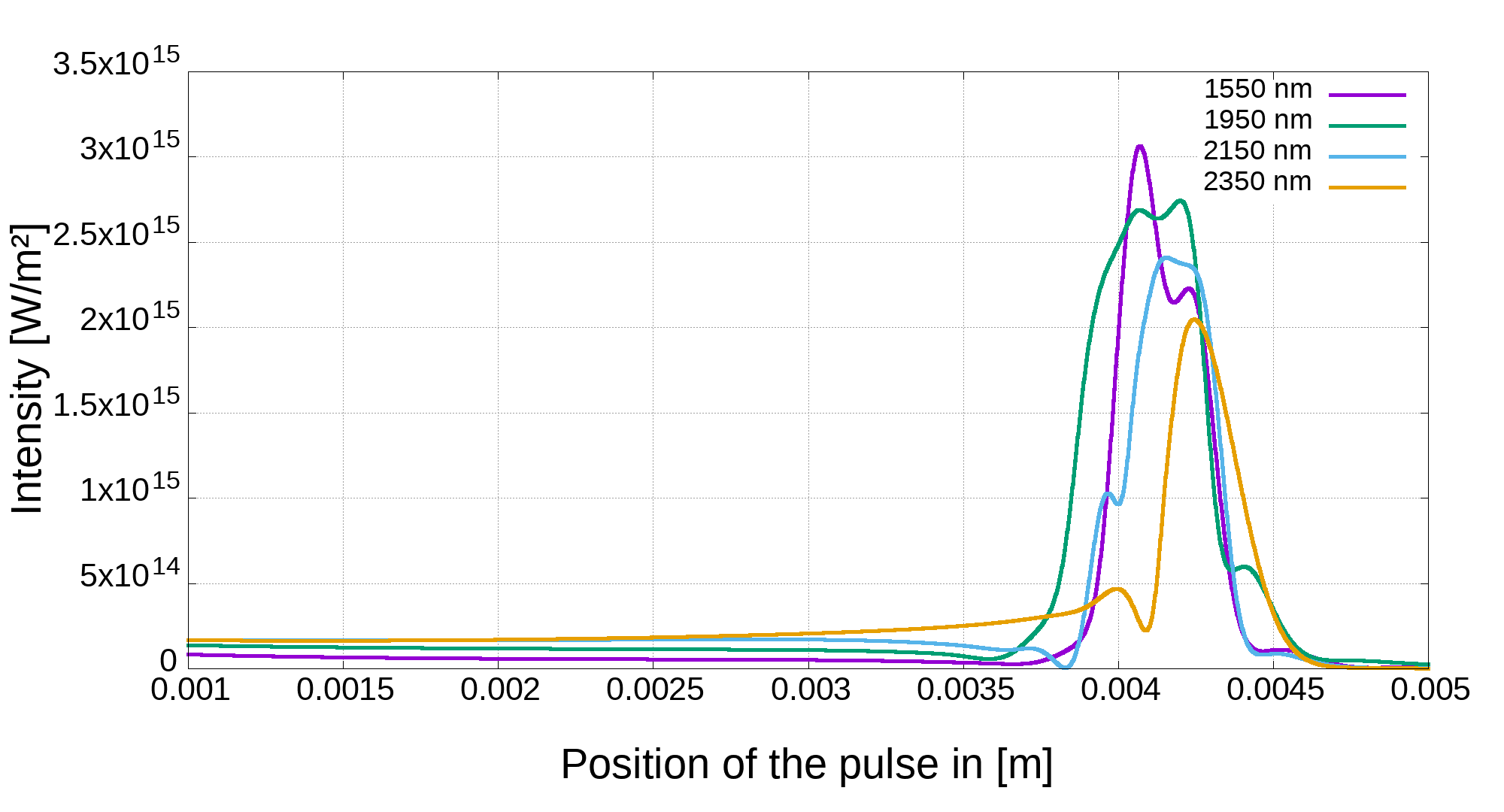}
    \caption{Low $n_2$-values, \SI{10}{\micro\joule} pulse energy}
    \label{fig:lown210u_int}
    \end{subfigure}
    \caption{Intensity of a pulse with different wavelengths and energies, depending on the position within the material. The focal length of the used lens is $f = \SI{8}{\milli\meter}$, with the focal point at $f_i = \SI{4}{\milli\meter}$ inside the material. The pulse duration is of $\SI{5}{\pico\second}$, and the pulse energy is varied between \SI{1}{\micro\joule} and \SI{10}{\micro\joule}.}
    \label{fig:intensity1u10u100u}
\end{figure}

\begin{figure}[htpb]
    \centering
    \begin{subfigure}[t]{\pythonimagescaling\textwidth}
    \includegraphics[width=\textwidth]{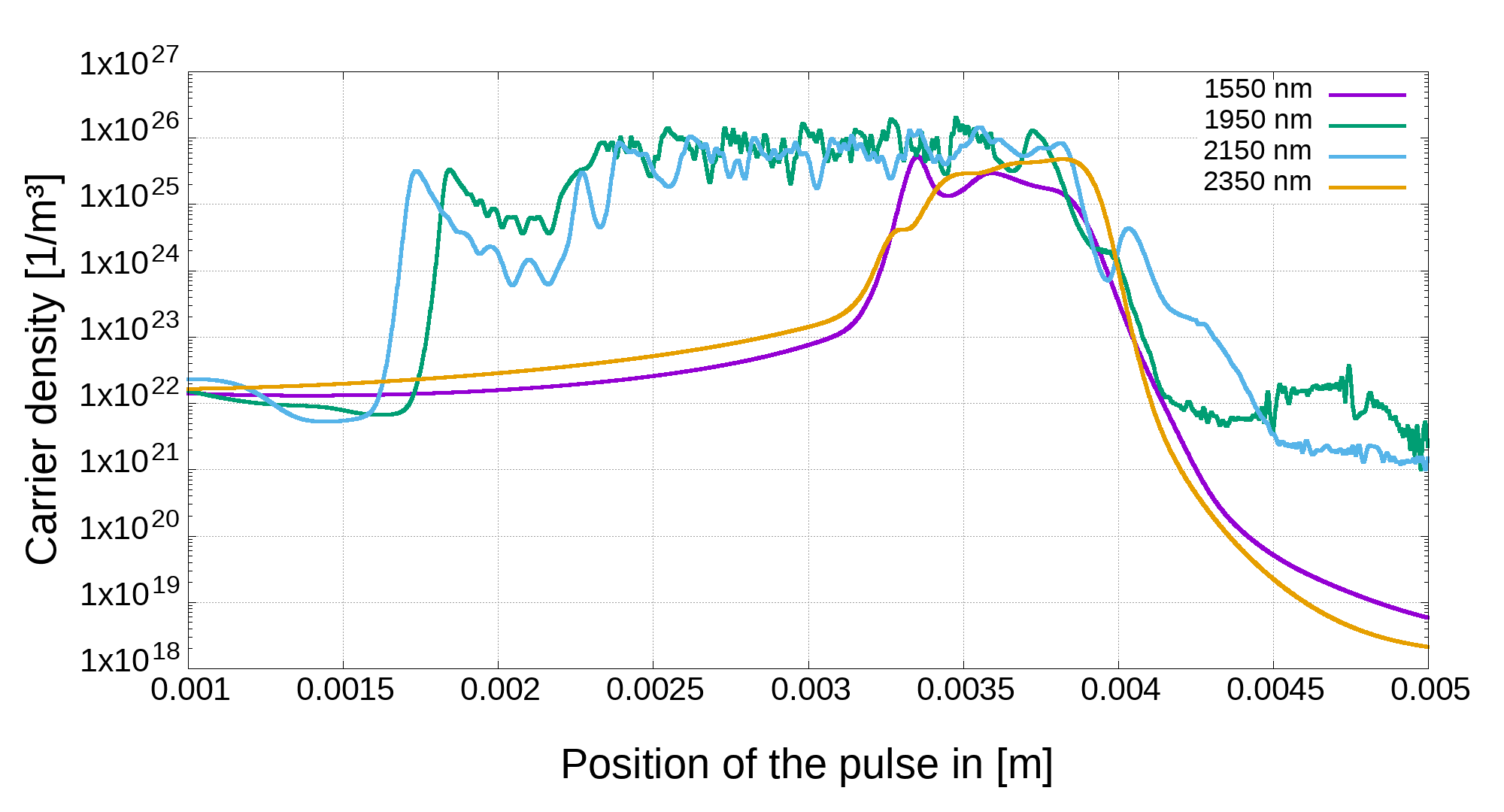}
    \caption{High $n_2$-values, \SI{1}{\micro\joule} pulse energy}
    \label{fig:highn21u_car}
    \end{subfigure}
    \hfill
    \begin{subfigure}[t]{\pythonimagescaling\textwidth}
    \includegraphics[width=\textwidth]{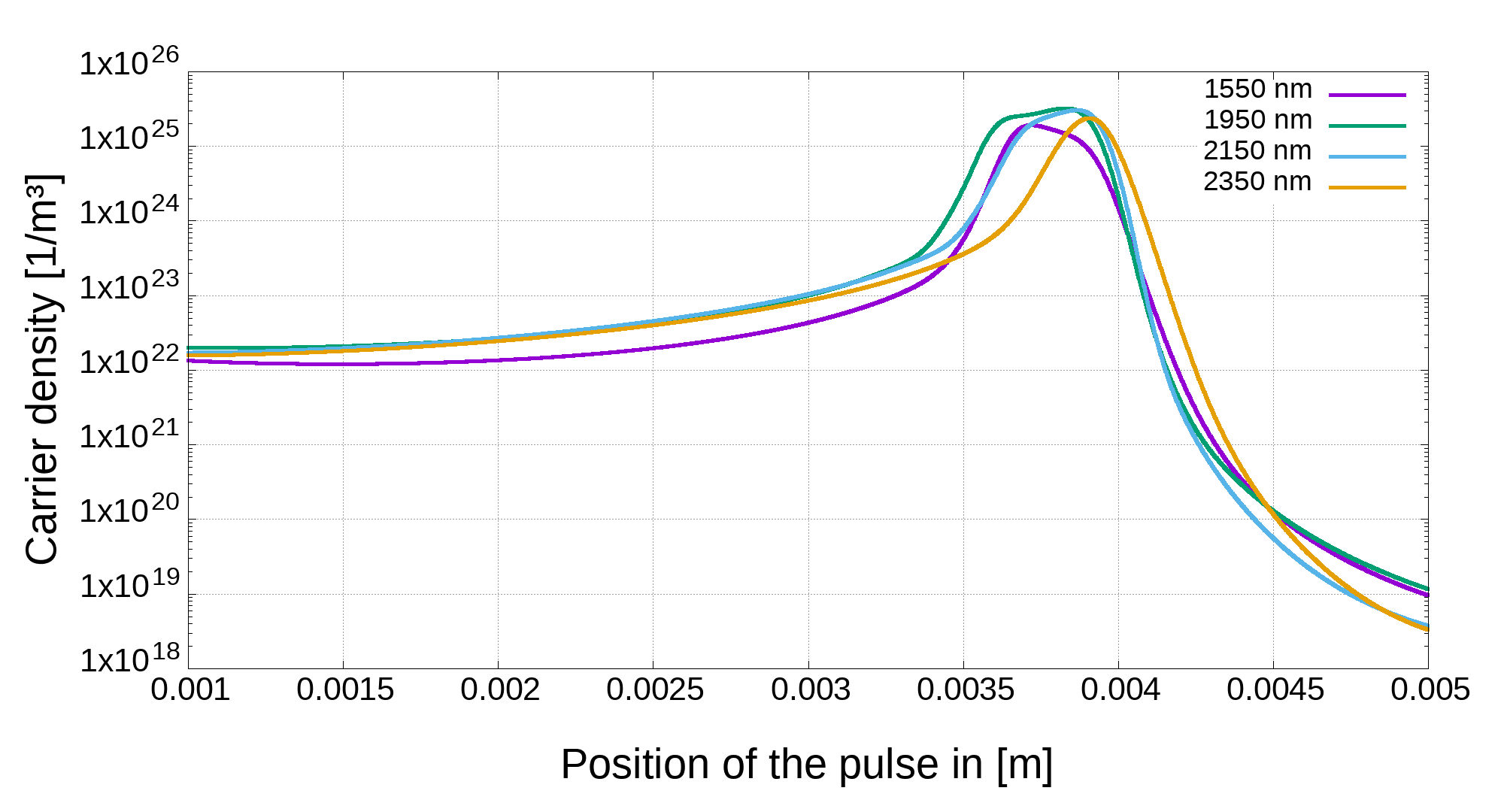}
    \caption{Low $n_2$-values, \SI{1}{\micro\joule} pulse energy}
    \label{fig:lown21u_car}
    \end{subfigure}
\hfill
    \begin{subfigure}[t]{\pythonimagescaling\textwidth}
    \includegraphics[width=\textwidth]{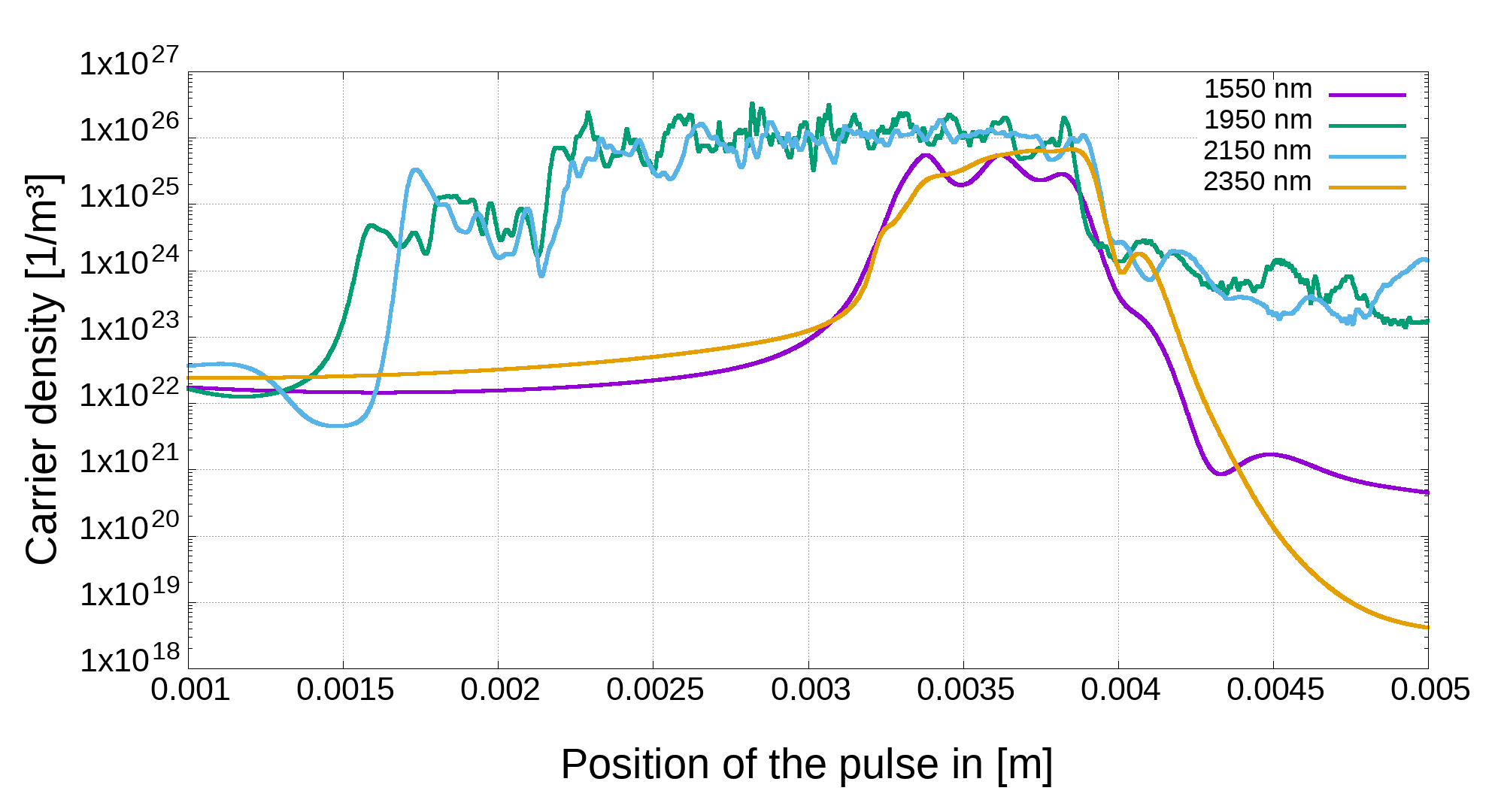}
    \caption{High $n_2$-values, \SI{10}{\micro\joule} pulse energy}
    \label{fig:highn210u_car}
    \end{subfigure}
\hfill
    \begin{subfigure}[t]{\pythonimagescaling\textwidth}
    \includegraphics[width=\textwidth]{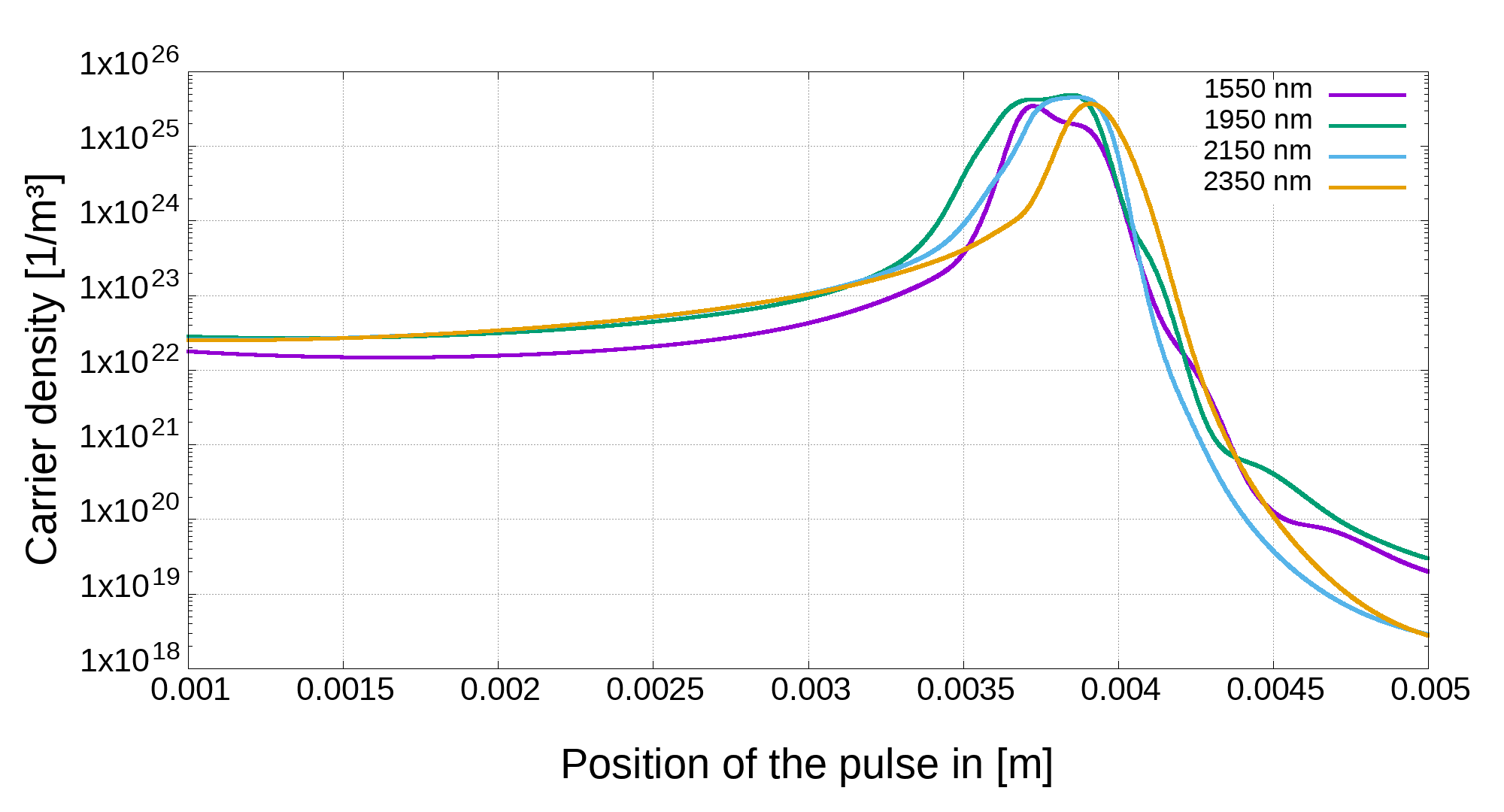}
    \caption{Low $n_2$-values, \SI{10}{\micro\joule} pulse energy}
    \label{fig:lown210u_car}
    \end{subfigure}
    \caption{Generated free carrier density from a pulse with different wavelengths, depending on the position within the material. The focal length of the used lens is $f = \SI{8}{\milli\meter}$, with the focal point at $f_i = \SI{4}{\milli\meter}$ inside the material. The pulse duration is $\SI{5}{\pico\second}$, and the pulse energy is varying between \SI{1}{\micro\joule} and \SI{10}{\micro\joule}.}
    \label{fig:carrier1u10u100u}
\end{figure}

While the pulse with the shortest wavelength is absorbed quite significantly already on the way to the focal spot, the other wavelengths can retain more energy, before depositing it closer to the focal spot. 
In addition, the absorption close to reaching the focal spot depends on the value of the third-order non-linearity, as shown e.g.~when comparing \Cref{fig:highn21u} and \Cref{fig:lown21u}. 
Here one also can see the ripples, originating from the strong self-focusing within the material, followed by local absorption and collapse, which show up in \Cref{fig:intensity1u10u100u,fig:carrier1u10u100u}, too. 

In the \Cref{fig:highn21u,fig:lown21u,fig:highn210u,fig:lown210u,fig:highn2100u,fig:lown2100u} most of the pulse energy is absorbed already in the first millimeter of the material, leading to a loss of \SIrange{50}{99}{\percent} of the pulse energy, depending on the initial pulse energy and wavelength. This process is independent of the value of the Kerr-nonlinearity. This high loss also means that the pulse has $ \approx\SI{0.5}{\micro\joule}$ left for \SI{1}{\micro\joule} initial pulse energy, $ \approx\SI{0.9}{\micro\joule}$ is left for \SI{10}{\micro\joule} initial pulse energy and $ \approx\SI{0.9}{\micro\joule}$ is left for \SI{100}{\micro\joule} initial pulse energy, which in turn shows that an increase of the pulse energy does not necessarily increase the energy transported to the focal spot, but instead increases the chance of damaging material on the way to the focal spot. 
Due to the similarity of the transported energy for the pulse with \SI{10}{\micro\joule} and with \SI{100}{\micro\joule} the profile for the generated intensity and carriers are very similar. Thus, those figures are omitted in \Cref{fig:intensity1u10u100u} and \Cref{fig:carrier1u10u100u}.

When considering the achieved intensities, one can see in \Cref{fig:intensity1u10u100u} both the effect of the nonlinear refractive index and the collapse of the pulse. 

In \Cref{fig:intensity1u10u100u} the self-focusing of the pulse at a wavelength of \SI{1950}{\nano\meter} and \SI{2150}{\nano\meter} starts at a significantly smaller depth than for the other two wavelengths. Self-focusing at a smaller depth leads to a modification zone which is spread out much more, compared to a pulse which is focused at a larger depth. 

The achieved intensity for the shortest wavelength (\SI{1550}{\nano\meter}) is higher than for other wavelengths when assuming the lowest reported values for the Kerr-effect in silicon, even though -- as it is shown in \Cref{fig:energy1u10u100u} -- a much lower portion of the pulse energy is transported to the focal spot for this wavelength. This can be explained by the interplay of stronger focusing at shorter wavelengths, something which is compensated for the longer wavelengths with an increasing Kerr-effect around \SIrange{2}{2.2}{\micro\meter} and thus even tighter focusing.

Finally, the effect already manifested in \Cref{fig:energy1u10u100u} also shows up in the intensity curves: Since most of the energy of the pulse is absorbed in the first small part of the beam path, the resulting energy at the focal spot is approximately the same for 10 and \SI{100}{\micro\joule}, being only slightly higher than that for \SI{1}{\micro\joule}. This results in the same intensity levels for the former two, and only slightly lower intensity levels for the latter one.
Higher intensity also means higher absorption, which can be calculated using \cite{Zavedeev2016}
\begin{equation}\label{equ:Absorption_Equation}
\partial_tI=\sum_K\beta^{(K)}I^K+\sigma N I
\end{equation}
Here, the absorption does not only depend on the non-linear absorption, but also on the free carrier density, which in turn is increased by the pulse itself (see \Cref{equ:CarrierDensityGeneration}). This generated free-carrier density is shown in \Cref{fig:carrier1u10u100u}, both for low and high $ n_2 $-values.

Again, the pulse at \SI{1950}{\nano\meter} and \SI{2150}{\nano\meter} can generate more free carriers, compared to the other two wavelengths, which in turn increases absorption and diffraction. This also holds true when considering the lower Kerr-nonlinearity. Even though the shortest wavelength could generate the highest intensity, the amount of the generated free carriers is approximately the same for all wavelengths, because of the higher free-carrier-generation efficiency at longer wavelengths. 

Moreover, because of the similar intensities for all pulse energies, the generated free carrier densities are also at approximately the same level, thereby confirming again that increasing the pulse energy from \SI{10}{\micro\joule} to \SI{100}{\micro\joule} does not increase modifications, but rather hinders them by destroying the material on the way to the focal spot.

When considering extended models, such as Two Temperature Models (TTM) as for example in \cite{Verburg2014} and \cite{Korfiatis2007}, the free carrier density also plays an important role by coupling the energy of the generated free carriers to the lattice itself by the coupling constant \cite{Verburg2014}
\begin{equation}\label{equ:CouplingConstant}
	\gamma(N)=\frac{3k_BN}{\tau_0\left(1+\left(\frac{N}{N_{crit}}\right)^2\right)}
\end{equation}
Thus, a higher free carrier density makes the energy transfer from the excited carriers to the lattice more efficient. This assumption must be still investigated further by coupling the pulse propagation to a two-temperature model. Finally, two different processes can lead to melting \cite{Perez2003,Zier2015}. On the one hand, the carrier density can be increased up to a critical carrier density \cite{Korfiatis2007,Sokolowski-Tinten2000}. At that density, enough free carriers are excited that the material can show non-thermal melting \cite{Chichkov1996,Zier2015,Shank1983,Stampfli1992}. This critical density ($ n_c\approx\SI{1e22}{\per\cubic\centi\meter}\equiv\SI{1e28}{\per\cubic\meter} $\cite{Sokolowski-Tinten2000}) is reached when approximately \SIrange{10}{20}{\percent} of the valence electrons are excited. This state shows a behaviour similar to melting, while the lattice temperature is still below the melting temperature. The other process is conventional thermal melting, when the lattice reaches a temperature above the melting temperature. 

Based on the numerical simulation data shown above (\Cref{fig:energy1u10u100u,fig:intensity1u10u100u,fig:carrier1u10u100u}) we can conclude that the created modifications in Si using the lens with $ f=\SI{8}{\milli\meter} $ and $ f_i=\SI{4}{\milli\meter} $ have a thermal nature, since the generated free-carrier density did not exceed the critical density. 

Our numerical simulations also confirm the initial assumption, that the wavelengths in the range of \SIrange{2000}{2200}{\nano\meter} are more efficient for modifying Si compared to wavelengths above \SI{2200}{\nano\meter} and below \SI{1900}{\nano\meter}. 

\section{EXPERIMENTAL}
\label{sec:ExpDiscussion}

\subsection{Laser sources}
\label{subsec:LasSourc}

In order to get experimental data which can be used as verification of the numerical results, modifications on and below the surface of Si-wafers were made using three different home-built lasers based on specialty fibers: An all-fiber LMA-based MOPA (Master Oscillator, Power Amplifier) system based on \ce{Tm} operating at a central wavelength of $ \lambda=\SI{2100}{\nano\meter} $, $ \tau_p=\SI{2}{\pico\second} $ and $ W_{max}=\SI{980}{\nano\joule} $ (see \Cref{subsubsec:2umFib}), a fibre-based \ce{Ho}-CPA laser operating at $ \lambda=\SI{2090}{\nano\meter} $, $ \tau_p=\SI{5}{\pico\second} $ and $ W_{max}=\SI{760}{\micro\joule} $ (see \Cref{subsubsec:SSL2090}) and an \ce{Er}-fiber based Cr:ZnS laser operating at $ \lambda=\SI{2350}{\nano\meter} $, $ \tau_p=\SI{1.7}{\pico\second} $ and $ W_{max}=\SI{20.3}{\nano\joule} $ (see \Cref{subsubsec:SSL2350}). 

\subsubsection{Tm-based fibre MOPA, operating at \texorpdfstring{$ \lambda=\SI{2}{\micro\meter} $}{l=\SI{2}{\micro\meter}}}
\label{subsubsec:2umFib}
The fibre-MOPA (labelled laser "A" in this paper) consists of an oscillator, mode-locked by a SESAM (Semiconductor Saturable Absorber), and two amplification stages. Here, the oscillator generates laser pulses at a wavelength of $ \lambda=\SI{1965}{\nano\meter} $ with a repetition rate of $ f=\SI{7.6}{\mega\hertz} $, a pulse duration of $ \tau_p=\SI{2}{\pico\second} $ and a pulse energy of $ W=\SI{0.1}{\nano\joule} $. By using a pre-amplifier this pulse energy is amplified to $ W=\SI{0.5}{\nano\joule} $. Finally, the pulse is amplified in a second amplifier to a pulse energy of $ W=\SI{980}{\nano\joule} $\cite{Richter:19}.

\subsubsection{Ho-fibre based laser at \texorpdfstring{$ \lambda =\SI{2090}{\nano\meter}$}{l=\SI{2090}{\nano\meter}}}
\label{subsubsec:SSL2090}
For experiments at $ \SI{2090}{\nano\meter} $ we used a compact ultra-short pulsed fibre laser based Ho:YAG amplifier system from ATLA Lasers AS. At \SI{10}{\kilo\hertz} repetition rate the system delivered up to \SI{760}{\micro\joule} pulses with \SI{5}{\pico\second} duration from the amplifier stage. This laser system was used both with \SI{500}{\femto\second}-pulses and \SI{5}{\pico\second}-pulses, but in this paper we present only the results from the longer \si{\pico\second}-pulses.

\subsubsection{Er-fibre based Cr:ZnS laser at \texorpdfstring{$ \lambda=\SI{2350}{\nano\meter} $}{l=\SI{2350}{\nano\meter}}}
\label{subsubsec:SSL2350}
The \ce{Er}-based fibre laser (labelled laser "C" in this paper) was a mode-locked Cr:ZnS oscillator from ATLA Lasers AS working in the normal dispersion regime \cite{Tolstik2018}, pumped by the \ce{Er}:fiber laser.
%
%
The combination of dispersion-managed mirrors and material dispersion compensation in the cavity ensured the normal-dispersion operation regime of the laser. The laser generated linearly chirped pulses centred at \SI{2350}{\nano\meter} with a duration of about $ \tau_p=\SI{1.7}{\pico\second} $ and pulse energy up to \SI{48}{\nano\joule} at a pulse repetition rate of \SI{3.8}{\mega\hertz}. Around \SI{50}{\percent} of the pulse energy was lost in the isolator, beam delivery optics and focusing objective, resulting in slightly above \SI{20}{\nano\joule} delivered to a surface of a silicon sample. 
%

\subsection{Experimental setup}
\label{subsec:ExpSetup}
All lasers were coupled to a processing stage. The light was focused using an aspheric lens with varying focal distances, typically $ f = \SI{4}{\milli\meter} $ either onto the surface of the sample (as in \Cref{fig:SurfaceLines}) or inside the sample (as e.g.~in \Cref{fig:vertical_lines_side_view}). The sample itself was mounted onto a four-axis stage, with two axes controlled by an electronic controller. Electronic control allowed precise movement with a fixed speed, variable between several \si{\micro\meter\per\second} and several \si{\milli\meter\per\second}. The sample itself was made out of mono-crystalline silicon.

\subsection{Investigation of the processed samples}
\label{subsec:InvestSamples}
The processed samples were investigated using three different techniques, including optical microscopy, transmission IR microscopy and scanning electron microscopy (SEM). The first two techniques are non-destructive, while the cross-sectional imaging using the SEM required splitting of the sample. 
Optical microscopy alone is applicable to study of the surface, while transmission IR microscopy was used to look through the sample and investigate sub-surface modifications without damaging the sample.
As mentioned in \Cref{sec:IntroSotA}, both surface and sub-surface silicon modifications were created using pulses at both \si{\nano\joule} and \si{\micro\joule} pulse energy levels. In the first part of this section (\Cref{subsubsec:SiModnJ}) the modifications created by pulses at sub-\si{\micro\joule} pulse energy levels are investigated. In the second part (\Cref{subsubsec:SiModuJ}) we investigate the modifications created by pulses with energies above \SI{1}{\micro\joule}.

\subsubsection{Silicon modifications created at \si{\nano\joule} pulse energy levels}
\label{subsubsec:SiModnJ}
We used the \ce{Tm}-fibre-MOPA (laser A), operating at $ f=\SI{7.6}{\mega\hertz} $ and $ W=\SI{28}{\nano\joule} $, and the \ce{Er}-based fibre laser, operating at $ \lambda=\SI{2350}{\nano\meter} $, $ f = \SI{3.7}{\mega\hertz} $ and $ W=\SI{17}{\nano\joule} $ (laser C), to create modifications both on the surface and below the surface at \si{\nano\joule} pulse energy levels. As initial test we created lines on the surface of the silicon sample, which directly could be observed with an optical microscope.
\begin{figure}[htpb]
    \centering
    \includegraphics[width=0.50\textwidth]{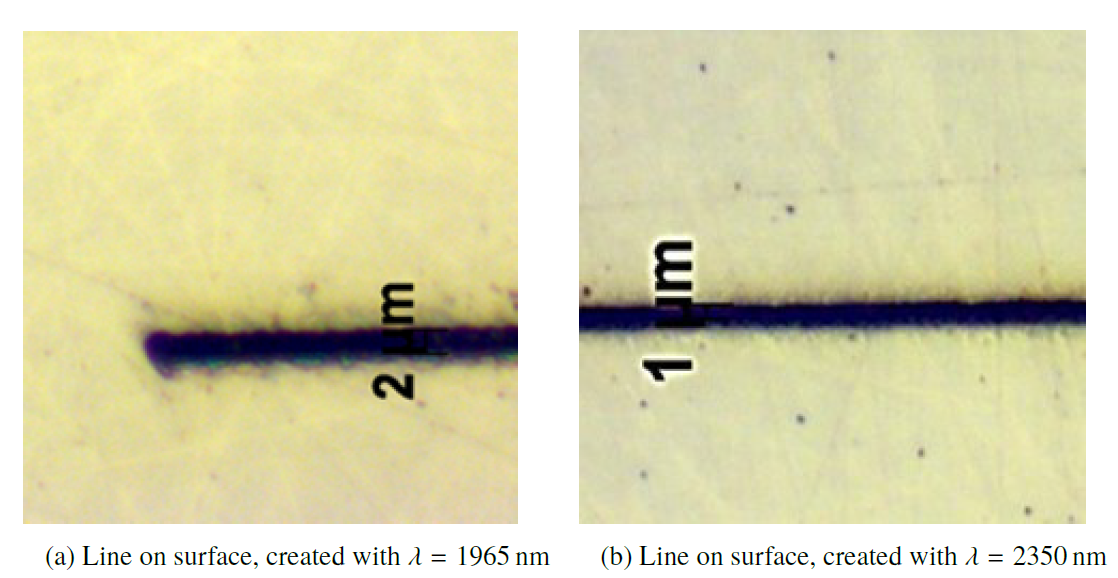}
    \caption{Lines on surface, created with $\lambda=\SI{1965}{\nano\meter}$ and $\lambda=\SI{2350}{\nano\meter}$ \cite{Richter:18}}
    \label{fig:SurfaceLines}
\end{figure}

For comparison in \Cref{fig:SurfaceLines} the cut using the longer wavelength is significantly smaller in diameter (\SI{1}{\micro\meter} compared to \SI{2}{\micro\meter}) and cleaner with less residual around the cut, thus suggesting that using a longer wavelength might be beneficial for cutting of Si. 

This size reduction could originate from a smaller spot size. Assuming that the beam was focused perfectly onto the surface, the focal spot size for laser A was \SI{4.25}{\micro\meter}, while for laser C the spot size was \SI{1.6}{\micro\meter}. Furthermore, we assumed that the absorption for laser A was purely based on two-photon-absorption, and the absorption from laser C was purely based on three-photon-absorption. This means that the effective spot size (if the threshold is reached for the whole Gaussian pulse) is \SI{3}{\micro\meter} for laser A, and \SI{0.92}{\micro\meter} for laser C. While the result for laser C is close to the measured width of the lines, the calculated effective spot size for laser A is significantly larger than the measured line width. This can be related to several things, such as an incorrectly measured beam width (the effective spot size can be reached for a beam width of \SI{1.6}{\milli\meter}), or a higher modification threshold, thereby also reducing the effective spot size. 
None of these possibilities can be excluded based on existing data. 

\begin{figure}[htpb]
    \centering
    \includegraphics[width=0.5\textwidth]{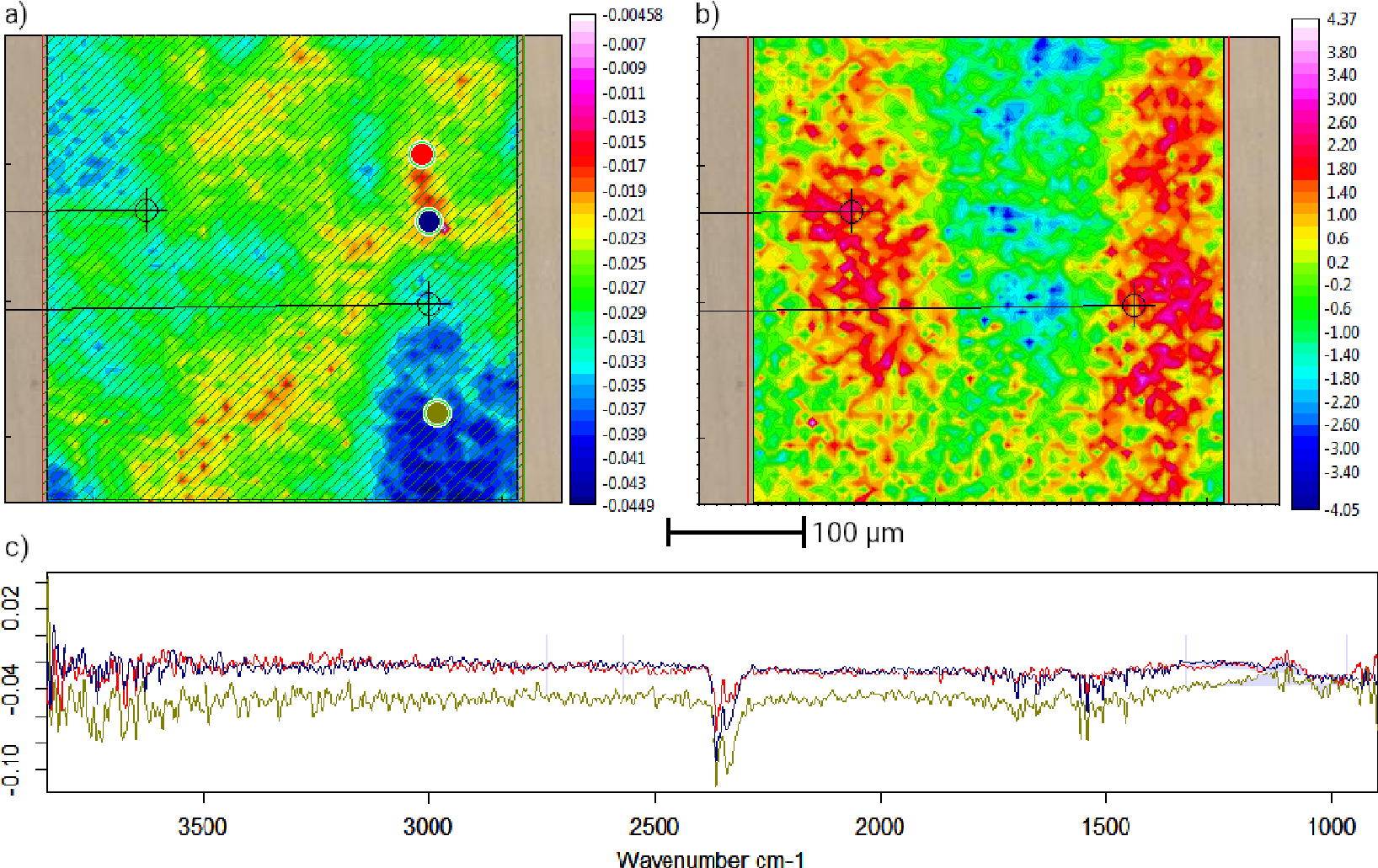}
    \caption{Transmission IR images of a line below the surface of Si  created with laser A operating at $\lambda=\SI{1965}{\nano\meter}$ and $W=\SI{180}{\nano\joule}$. a) Mapped integrated extinction in the range \SIrange{3.965}{3.027}{\micro\meter}; b) mapped integrated absorbance in the region \SIrange{10.320}{7.553}{\micro\meter}. The scale bar applies to both a) and b). c) Example spectra at the spots marked in the respective colour in a). As in \Cref{fig:ftir_single_image}, the continuation of the sample is shown as a visible image outside the IR maps.}
    \label{fig:two_images}
\end{figure}

\begin{figure}[htpb]
    \centering
    \includegraphics[width=0.35\textwidth]{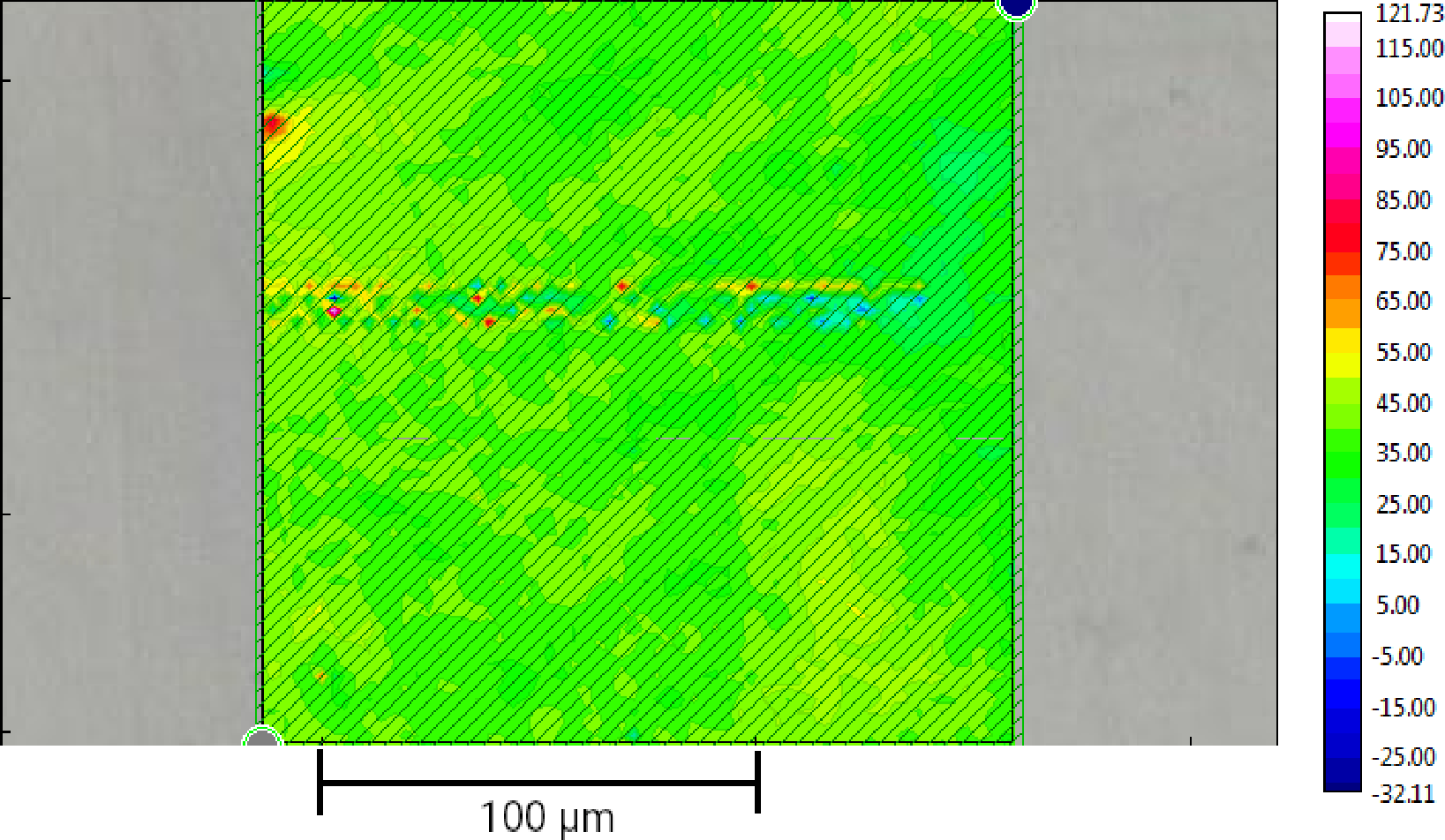}
    \caption{Transmission IR image of a line created by focusing laser C with $\lambda=\SI{2350}{\nano\meter} $ below the surface of silicon. Right: False colored absorbance scale in the wavelength range \SIrange{3.965}{3.027}{\micro\meter}. The visible part of the image left and right of the central false coloured image presents a continuation in the visible of the area imaged in the IR. No modification is visible there, thus indicating that the modification resulting in increased absorption only exists below the surface.}
    \label{fig:ftir_single_image}
\end{figure}

After we could confirm that we can create modifications at the surface of silicon, we have conducted also the sub-surface modifications in silicon. For this task the same lasers as mentioned above were used, with varying pulse energy levels. Those modifications were observed both using a transmission FTIR and an infrared microscope. The FTIR microscopy was carried out in transmission mode on a Bruker Hyperion 3000 microscope with a $64\times64$ pixel mercury-doped \ce{CdTe} focal plane array detector interfaced to a Bruker Vertex 70v FTIR spectrometer. A $ 10\times $ Cassegrain objective was used for imaging. Background spectral images were recorded on an empty sample stage and used to calculate extinction spectra at each pixel after measuring the laser processed Si sample. For spectral acquisition, 100 scans with a spectral resolution of $ \SI{4}{\per\centi\meter} $ were averaged. Spectral images were obtained from two spectral regions. Extinction maps from the region $ \SIrange{3.965}{3.027}{\micro\meter} (\SIrange{2522}{3304}{\per\centi\meter}) $ were obtained as false-coloured maps from integrating extinction without baseline correction. As the system does not show any absorption in this spectral range, extinction here is caused by scattering. Maps were also produced in the spectral region \SIrange{10.320}{7.553}{\micro\meter} (\SIrange{969}{1324}{\per\centi\meter}), containing the absorption peaks of Si-O stretching modes. In this region, a linear baseline was subtracted from the spectra between the left and the right end of the spectral range. In that manner, maps show the distribution of absorption from Si-O modes. Results of those measurements for the laser operating at \SI{1965}{\nano\meter} (laser A)
 and for the laser operating at \SI{2350}{\nano\meter} (laser C) are displayed in \Cref{fig:two_images,fig:ftir_single_image}. 

\subsubsection{Silicon modifications created at \si{\micro\joule} pulse energy levels}
\label{subsubsec:SiModuJ}
\begin{figure}[htpb]
    \centering
    \begin{subfigure}[t]{0.49\columnwidth}
    \centering
        \includegraphics[width=0.9\linewidth]{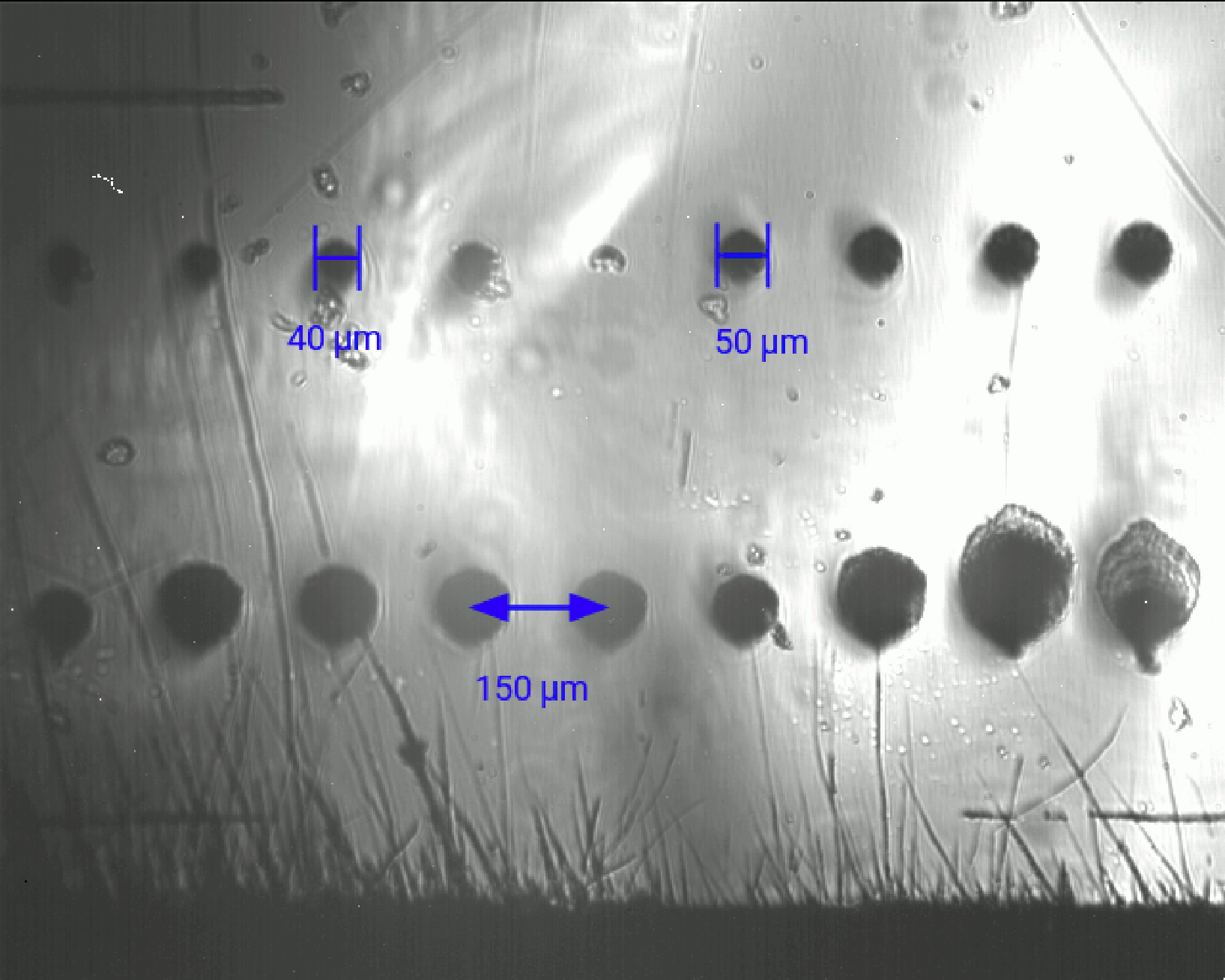}
        \caption{View from the top}
        \label{fig:vertical_lines_top_view}
    \end{subfigure}%
    \begin{subfigure}[t]{0.49\columnwidth}
    \centering
        \includegraphics[width=0.9\linewidth]{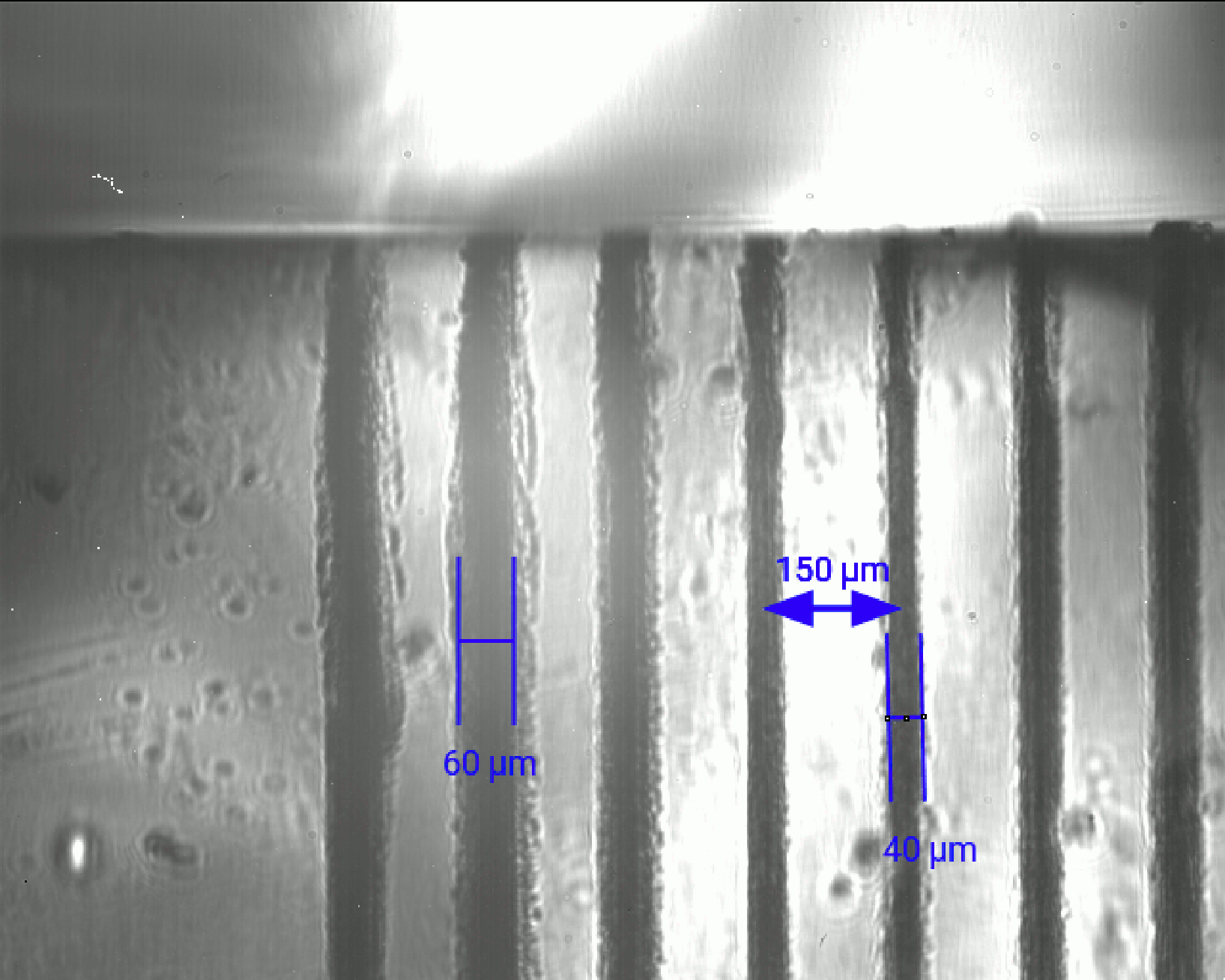}
        \caption{View from the side}
        \label{fig:vertical_lines_side_view}
    \end{subfigure}
    \caption{Vertical lines, viewed using a transmission IR microscope from the top of the sample (\Cref{fig:vertical_lines_top_view}) and the side of the sample (\Cref{fig:vertical_lines_side_view}). The pulse energy was \SI{3.7}{\micro\joule} and \SI{7.7}{\micro\joule}. The movement speed of the stage decreases from left to right.}
    \label{fig:vertical_lines_in_SI}
\end{figure}

By using laser B operating at $\lambda=\SI{2090}{\nano\meter}$, we increased the pulse energy to several \si{\micro\joule}. We used this increased pulse energy for creating larger sub-surface modifications, and investigated their shape afterwards. In addition, for faster non-destructive detection of sub-surface modifications in bulk silicon (compared to the FTIR used in \Cref{subsubsec:SiModnJ}) we designed a special transmission IR-microscope operating at $ \lambda=\SI{1300}{\nano\meter} $.
To increase the modifications' visibility, the sample was moved parallel to the beam, while the samples shown in \Cref{fig:two_images,fig:ftir_single_image} were moved perpendicular to the laser beam. This movement resulted in visible spots at the surface of the sample, shown in \Cref{fig:vertical_lines_top_view}.

The focal spot within the sample was moved vertically with a speed between $ \SIrange{0.2}{20}{\micro\meter\per\second} $ from the bottom of the sample towards the top surface, while the pulse energy was varied between \SIrange{3.7}{7.7}{\micro\joule}. A side view of a part of the sample is shown in \Cref{fig:vertical_lines_side_view}. 

\begin{figure}[htpb]
    \centering
    \begin{subfigure}[t]{0.49\columnwidth}
    \centering
        \includegraphics[width=0.99\linewidth]{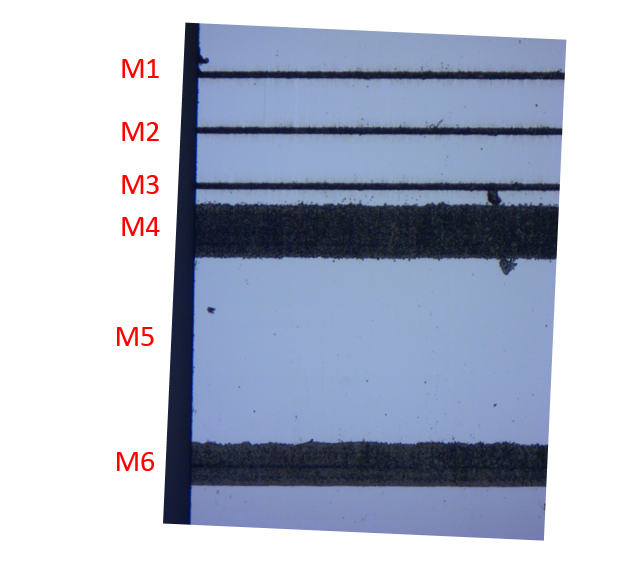}
        \caption{View using the optical microscope}
        \label{fig:top_view_surface_opt}
    \end{subfigure}%
    \hfill
    \begin{subfigure}[t]{0.49\columnwidth}
    \centering
        \includegraphics[width=0.99\linewidth]{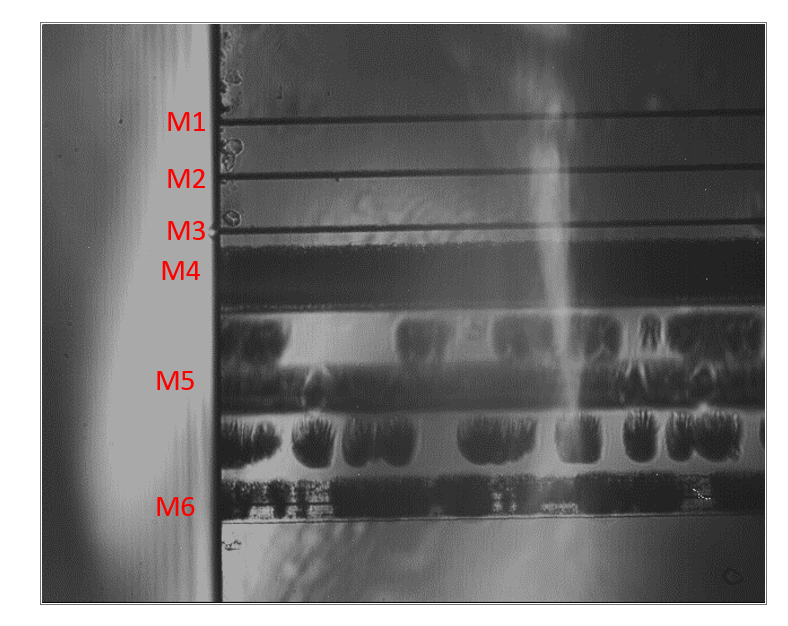}
        \caption{View using the infrared microscope}
        \label{fig:top_view_surface_ir}
    \end{subfigure}
    \caption{View of the surface of the sample, using the optical microscope, and through the sample, using an infrared microscope.}
    \label{fig:surface_dual_view}
\end{figure}

\begin{figure}[htpb]
    \centering
    \includegraphics[width=0.2\textwidth,angle = 90]{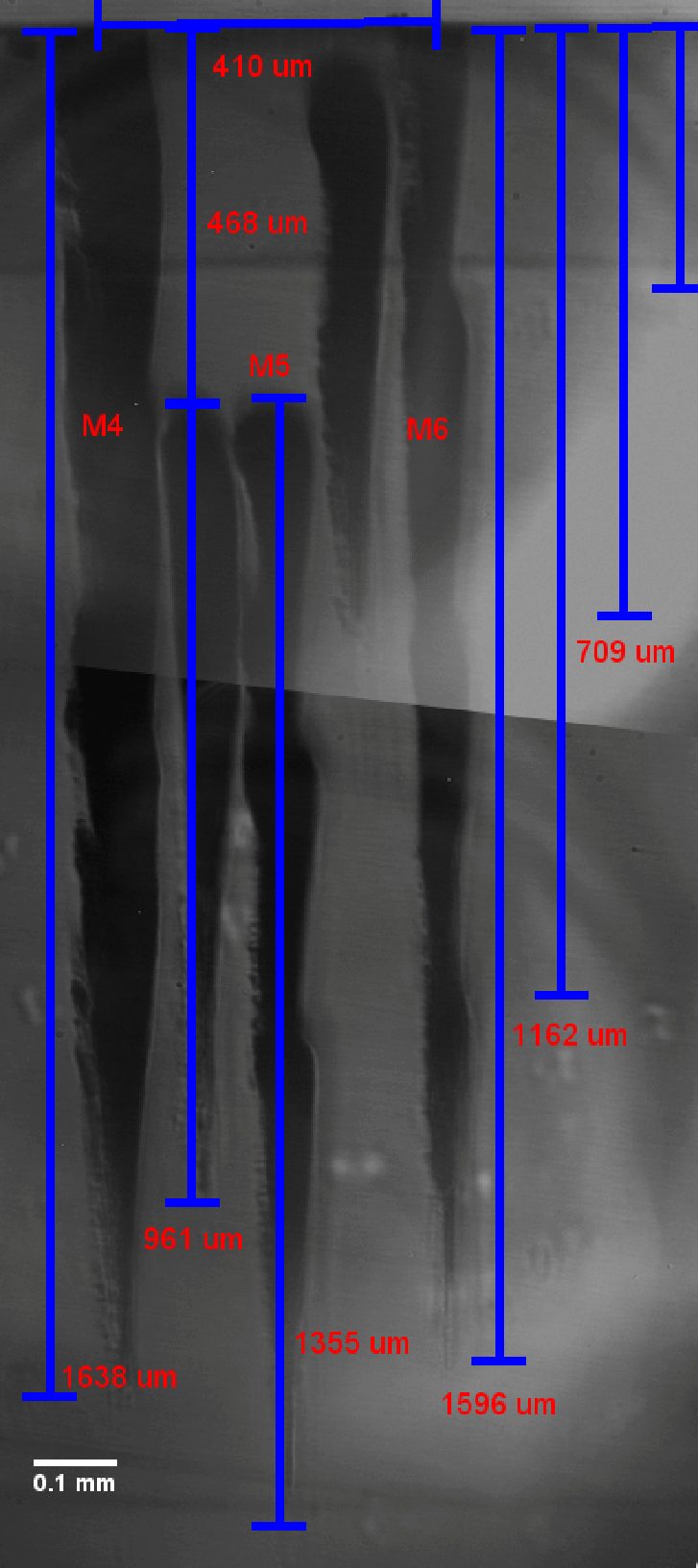}
    \caption{Horizontal lines created in Si, watched from the side. Line stacking clearly is visible}
    \label{fig:horizontal_lines_ir_view}
\end{figure}

\begin{figure}[htpb]
    \centering
    \includegraphics[width=0.45\textwidth]{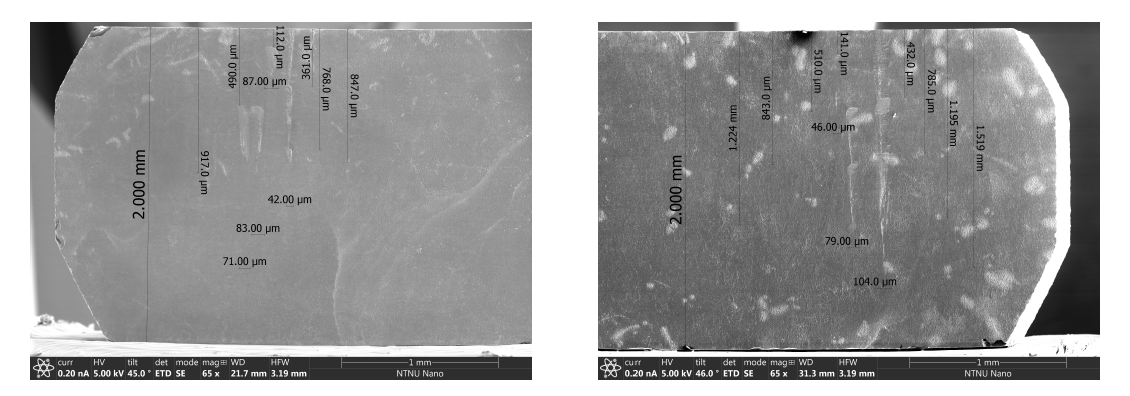}
    \caption{SEM image of the sample surface. The left figure shows the front side of the sample, the right figure the back side of the sample}
    \label{fig:SurfaceSEMView}
\end{figure}

A second sample translated perpendicular to the beam using the same parameters is shown in \Cref{fig:surface_dual_view,fig:horizontal_lines_ir_view,fig:SurfaceSEMView}. Each line is marked, with the marks in \Cref{fig:surface_dual_view,fig:horizontal_lines_ir_view} and \Cref{fig:SurfaceSEMView} corresponding to each other.

There are modifications below the surface between line M4 and M6, even though they are not continuous. It also must be noted that lines M4, M5 and M6 are stacked lines, i.e.~several lines were drawn on top of each other. This "line-stacking" can also be seen when rotating the sample and watching it from the side using the IR microscope, as shown in \Cref{fig:horizontal_lines_ir_view}. After confirming that modifications below the surface were made, the sample was cleaved, and the cross section was investigated by SEM. The sample had to be observed from both sides, due to the spottiness of the lines. The different views from both sides are shown in \Cref{fig:SurfaceSEMView}. It clearly can be seen that only a part of the observed modifications are visible on each side of the sample, as for example line M6 from \Cref{fig:horizontal_lines_ir_view}. This line is only visible on the front side, but not on the back side. On the contrary, line M5 is visible on both sides. 

When comparing \Cref{fig:horizontal_lines_ir_view} with \Cref{fig:SurfaceSEMView}, it can be also noted that the depths of the modifications are approximately equal, thereby indicating that the modifications observed in \Cref{fig:horizontal_lines_ir_view} are the same as in \Cref{fig:SurfaceSEMView}, even though the modifications visible in \Cref{fig:horizontal_lines_ir_view} are larger than the modifications seen in \Cref{fig:SurfaceSEMView}. Another fact worth mentioning is the shape of the modifications. Even though the beam was focused to a certain depth, the modifications started already significantly earlier, leading to elongated shapes. This broadening will be discussed in \Cref{sec:Discussion}. 
\subsection{Summary}
\label{subsec:Summary}
We showed three different lasers operating at central wavelengths of \SIrange{1965}{2350}{\nano\meter} and two different levels of pulse energies (\si{\nano\joule} and \si{\micro\joule}) and the resulting modifications of silicon both on the surface and below.
The modifications were investigated in some more detail for $ \lambda=\SI{2090}{\nano\meter} $, which we identified as a "sweet spot" for silicon modifications. The sub-surface modifications we generated with this laser could be visualised both by a custom-built transmission IR microscope and a SEM. The structures visible with both methods correlated in size. 
Sub-wavelength feature size was evidenced in \Cref{fig:SurfaceLines}, but could not be observed yet for sub-surface structures as in \Cref{fig:vertical_lines_in_SI,fig:surface_dual_view,fig:horizontal_lines_ir_view,fig:SurfaceSEMView}. A possible reason for that is the nonlinear absorption, which is starting already before the focal spot, thereby creating modifications with sizes larger than the laser wavelength as shown in the numerical simulation in \Cref{fig:intensity1u10u100u}, and in the experiment in \Cref{fig:vertical_lines_in_SI,fig:surface_dual_view,fig:horizontal_lines_ir_view,fig:SurfaceSEMView}.

\section{DISCUSSION}
\label{sec:Discussion}
As it can be seen in \Cref{fig:horizontal_lines_ir_view} and other figures the created shapes within Si are not perfectly shaped around the focal spot, but rather elongated and start significantly before the focal spot. One possible explanation for that is that nonlinear absorption happens already earlier, i.e. the critical intensity for material modification is reached already before the focal spot, starting the modification process before reaching the focal point. A similar behaviour is shown in \Cref{fig:intensity1u10u100u}. One would expect the intensity to rise to one single peak, but instead it hits a maximum value, before staying at that level for several micrometres and decreasing again afterwards. In that period the pulse energy is absorbed within the material, and, depending on the pulse energy, is leading to modifications at points which are not targeted. This modified area also depends on the value for the non-linear Kerr value (as it can be seen in \Cref{fig:highn21u_int} compared to \Cref{fig:lown21u_int} for \SI{1}{\micro\joule}, and similarly in \Cref{fig:highn210u_int} and \Cref{fig:lown210u_int} for \SI{10}{\micro\joule}). The exact length of the modified area does not only depend on the pulse energy, but also on the value of the Kerr-coefficient. 

This also means that a higher Kerr coefficient allows processing of Si at lower pulse energies. Lower energies will result in lower absorption on the way to the focal spot, while the self-focusing will occur later, and thereby resulting in a smaller modified area, which in turn explains the sub-surface modifications shown in \Cref{fig:two_images} and \Cref{fig:ftir_single_image}. Both modifications were made at power levels significantly below \SI{1}{\micro\joule}, and thereby unable to damage the surface, while still modifying material below the surface.

Similar results are also reported by \cite{Chambonneau2019}, which reported a lower threshold of \SI{300}{\nano\joule} for creating sub-surface modifications in Silicon when applying multiple pulses at the same spot. Finally, their reported fluence distribution matches the shape of the lines showed in \cref{fig:horizontal_lines_ir_view}, which thereby confirms their simulations with experimental data.
\section{CONCLUSIONS}
\label{sec:Summary_and_outlook}
In this paper we investigated how the wavelength of the used laser influences the generated modifications at the surface and within silicon. This was done in three separate steps. On the one hand, we used a simple numerical model which included the wavelength dependency of the non-linear refractive index and the multi-photon absorption values, in \Cref{sec:NumSimulations}. On the other hand, we investigated the generation of modifications at two different pulse energy levels, at \si{\nano\joule} and at \si{\micro\joule} levels, with varying wavelengths.

As main conclusion of our results we can say that there is an optimal wavelength around \SI{2.1}{\micro\meter} where subsurface processing of Si occurs at comparatively low pulse energies.

Moreover, a few additional conclusions can be drawn. For example, for certain illumination parameters no visible damage on the surface was observed, but IR images showed a contrast in the volume of the material. Other groups experienced difficulties producing modifications inside the silicon volume without damaging the surface \cite{Sreenivas2012}. This knowledge can be used not only for creating a weakened layer below the surface, but also for creating wave guides. When comparing the pulse parameters given in \cite{Izawa2008,Tokel2017,Pavlov2017,Sreenivas2012,Kammer2018,Nejadmalayeri2005}, sub-surface modifications here were created at comparable or lower pulse energies. Consequently, modifications at longer wavelengths $ \geq\SI{1550}{\nano\meter} $, especially between \SI{2000}{\nano\meter} and \SI{2200}{\nano\meter}, have a lower damage threshold than at shorter and longer wavelengths, as predicted in \Cref{sec:NumSimulations}, and also confirmed in \cite{Chambonneau2019}. 

An increase of the pulse energy does not necessarily improve the processing; in some cases, it rather worsen the quality of the written structure by increasing the area where material is modified, though this could also be related to being the first results, which will need further improvement. In future studies we plan to apply a "burst pulse" approach and see if we can achieve improvements at both higher repetition rates and higher pulse energies per spot. 

As an outlook, due to the non-linear behaviour of multi-photon absorption, it is possible to create structures with a characteristic size below the diffraction limit, something which is not easily achievable through direct writing with single-photon absorption. 
Additionally, it was possible with this method to create modified 2d-layers buried within the bulk silicon, something which, to the best of our knowledge, has not been reported before. This will be the subject of an upcoming publication.


The exact nature of the modified Si must be investigated in the future in more detail on the physical-chemical level in order to gain a better insight, and to improve the process. Still, the already achieved first results are quite promising, and further research has to be done at the wavelengths above 2 microns.

\section*{ACKNOWLEDGEMENTS}
We would like to thank Prof. Patrick Espy very much for proofreading this work. This work was supported by ATLA Lasers AS, the Research Council of Norway –- ENERGIX project \#255003, as well as the Austrian Science Fund project \#P24916 and COST Action MP1401.  V.L.K. acknowledges the support by the Marie S.-Curie Cofund Multiply ``MASTEDIS'' Fellowship.





\bibliographystyle{IEEEtran}
\bibliography{IEEEabrv,bibliography.bib}
\end{document}